\documentclass[preprint,aps,12pt,showpacs,nofootinbib,tightenlines,amsmath,amssymb]{revtex4}
\usepackage{amsmath}
\usepackage{graphicx} 
\usepackage{amssymb}
\usepackage{color}

\newcommand{\p}{\partial}

\newcommand{\Tr}{\mathop{\rm Tr}\nolimits}

\begin{document}
\def\intdk{\int\frac{d^4k}{(2\pi)^4}}
\def\sla{\hspace{-0.17cm}\slash}
\hfill

%\begin{titlepage}
\title{Thermal Mass Spectra of Vector and Axial-Vector Mesons in Predictive Soft-Wall AdS/QCD Model}

\author{Ling-Xiao Cui}\email{clxyx@itp.ac.cn}

\author{Shingo Takeuchi}\email{shingo@itp.ac.cn}

\author{Yue-Liang Wu}\email{ylwu@itp.ac.cn}

\affiliation{State Key Laboratory of Theoretical Physics(SKLTP)\\
Kavli Institute for Theoretical Physics China (KITPC)\\
Institute of Theoretical Physics, Chinese Academy of Science, Beijing,100190, China}

\date{\today}

\begin{abstract}
We extend the predictive soft-wall AdS/QCD model to a thermodynamic model by considering the black hole metric. A modified bulk vacuum expectation value and a modified coefficient for the quartic term in the bulk scalar potential are introduced to obtain a smooth dilaton solution for computing the spectral functions of the vector and axial-vector mesons. It is demonstrated that the peaks appearing in the spectral functions characterize the thermal mass spectra of vector and axial-vector mesons,  where the location of the peak moves to a lower value and the width of the peak becomes wider when increasing the temperature. We observe that the peak disappears completely at the critical temperature around $T_c=200$ MeV, which implies the deconfinement of quark and the restoration of chiral symmetry breaking. A numerical study by fitting the spectral function in
terms of the Breit-Wigner form has been made to show how the peak dissolves quantitatively when the temperature is increased to the critical point.
%-----
\end{abstract}

\maketitle

\section{Introduction}
\label{Chap:Intro}

The reproduction of the extremely early moment after the big-bang is one of the biggest
purposes and achievements in the present high energy physics. The experimental exploration is expected to be carried out by the groups at LHC and RHIC, as well as many other groups including FAIR, J-PARC, JLab and RIBF. It will involve with the following basic problems:
the novel matter phase of quark-gluon plasma, confinement of colors and quarks, chiral symmetry breaking and restoration,
unstable nuclides, inner structure of hadrons and the force between hadrons, color glass condensation and superconductor. Unlike the perturbative QCD which is well understood based on the property of asymptotic freedom of QCD\cite{Gross:1973id,Politzer:1973fx}, all the studies will lead us to understand deeply the properties of nonperturbative QCD as they all are relevant to the strongly coupled QCD.
So far, the strongly coupled QCD is mainly described by the chiral dynamical models of QCD with dynamically spontaneous symmetry breaking\cite{Nambu:1960xd,DW,HW}, lattice QCD and holographic QCD (or AdS/QCD)\cite{AdSQCD}.

The holographic QCD has been developed based on the AdS/CFT duality to characterize the strongly coupled QCD.  AdS/CFT was initiated as the duality between the weakly coupled type IIB supergravity on $AdS_5 \times S^5$ space-time and
the strongly coupled $D=4$ ${\cal N}=4$ super Yang-Mills theory \cite{AdSCFT}.
It is known that $D=4$ ${\cal N}=4$ super Yang-Mills theory has conformal symmetry
and no field associated with fundamental representation. Thus AdS/CFT itself cannot correspond to the realistic QCD.
The attempt for making AdS/CFT approach to the realistic QCD was first pursued in ref.\cite{AdSQCD},
where the study has been paid to construct phenomenologically the model in searching for the bulk gravity matching with the realistic QCD.
Such kind of bottom-up approach is usually called as holographic QCD or AdS/QCD models which include hard-wall AdS/QCD model \cite{Erlich:2005qh,DP,Ghoroku:2005vt,SWWX} and soft-wall AdS/QCD model \cite{Karch:2006pv,Colangelo:2008us,Gherghetta:2009ac,pADSQCD1,pADSQCD2}. It is also interesting to observe the correspondence between matrix elements obtained in AdS/CFT with the corresponding formula
using the light-front representation as shown in Refs. \cite{BT1,BT2,BT3}.
It may distinguish from the top-down model resulted from solutions of string theories like D3/D7 \cite{D3D7_a}, D4/D8/$\overline{\rm D8}$ \cite{D4D8D8} and D4/D6/$\overline{\rm D6}$ \cite{D4D6}.

In AdS/QCD models, the cutoff is put artificially around the horizon in the radial coordinate of the bulk gravity.
The location of such a cutoff corresponds to the inverse of QCD scale $1/\Lambda_{QCD}$ in the dual field theories.
Here the radial coordinate in the bulk corresponds to the energy scale in the dual field theories, thus the cutoff has a role as the IR cutoff in the dual field theories. There is a difference in how the cutoff is put between hard-wall and soft-wall AdS/QCD model.
In the hard-wall model, the range of the radial coordinate is restricted sharply at the cutoff in the integration of Lagrangian.
Whereas in the soft-wall model, the action begin to be suppressed gradually from around $1/\Lambda_{QCD}$ to the horizon by introducing the dilaton which is regarded as a background field.

One of the main topics
in the holographic AdS/QCD is
to consistently incorporate both the dynamics of chiral symmetry
and the linear dependence of squared mass of mesons
on spin and excitation number,
namely the dynamically spontaneous chiral symmetry breaking
and the linear confinement of QCD.
The hard-wall model can succeed in the realization of the linear dependence
on lower excited states and lower spins,
while the linear dependence deviates at higher excited states or higher spins \cite{criticism}.
The soft-wall model was motivated
to improve this situation \cite{Karch:2006pv},
so that the linear dependence was found to be satisfied
for higher excited states and spins,
but the resulting chiral condensation is proportional to quark mass, which is inconsistent with the realistic QCD.
To further improve the situation, a higher order quartic term
was introduced in the potential of the bulk scalar field,
which modified the proportional
between chiral condensation and quark mass,
but the bulk scalar potential field is not bounded \cite{Gherghetta:2009ac}.
%----

It has been shown in ref.\cite{pADSQCD1,pADSQCD2} that
by simply modifying the bulk gravity at the infrared (IR) region
in the soft-wall AdS/QCD and keeping the conformal invariance
to be unchanged at the UV region as required from the property of QCD,
it enables us to build a predictive AdS/QCD model
which can provide a consistent prediction for the mass spectra of all light mesons.
As a consequence, the mass spectra for both the groundstate mesons and resonance mesons
can match well with the experimental data.

%\textcolor{red}{
Another interesting topics in the holographic AdS/QCD are
to investigate the finite temperature effects.
%---
In ref.\cite{Herzog:2006ra},
in both of the hard- and soft-wall models,
confinement/deconfinement transition temperatures
were computed based on the evaluation of the free energies.
%---
Ref.\cite{mssp_hard} is a study of meson in the hard-wall model at finite temperature.
%---
The finite temperature effect
on the spectrum of glueballs or mesons in the soft-wall model
was studied in ref.\cite{Colangelo:2009ra,Misumi,Miranda:2009uw}.
%---
In Ref.\cite{Grigoryan:2010pj},
a holographic model for charmonium was built
in the soft-wall model.
%---
In refs.
%\cite{mssp_soft,mssp_soft_ChiralBreaking},
\cite{Colangelo:2008us,mssp_soft_ChiralBreaking},
both mesons and gluons were investigated in the soft-wall model.
%---
In ref.\cite{Mas:2008jz,Erdmenger:2007ja,mspf_D3D7_0b},
the spectral function was computed
by taking the D3/D7 setup which is dual to
${\cal N}=4$ Super Yang-Mills theory
with fundamental matters
at finite baryon density.
%---
In ref.\cite{mspf_D3D7_E},
the spectral function was computed holographically
by modifying the D3/D7 setup
with finite baryon density \cite{Mas:2008jz,Erdmenger:2007ja,mspf_D3D7_0b}
such that the dual field theory can have electric background.
%---
The setup with D4 and $\overline{\rm D4}$-branes on $AdS_6$ was considered in ref.\cite{mspf_NCS}.
%---
This is a non-critical string setup,
which is considered
not to have the problems
originated from in critical string setups
like $D4/\overline{\rm D4}/D8$.
%---
In such a circumstance,
the spectral function was computed (hydrodynamics was also performed).
%}

In this paper,
we are going to extend the predictive soft-wall AdS/QCD model \cite{pADSQCD1,pADSQCD2}
to a thermodynamic model at finite temperature.
In particular, we will examine finite temperature effect on mass spectra of vector and axial-vector mesons
from the computation of the spectral function.
The paper is organized as follows:
In Sec.\ref{Chap:Model},
we include the finite temperature effect into
the predictive soft-wall AdS/QCD model
in the usual way
by simply considering a black hole metric.
In sec.\ref{Chap:VM},
we present a numerical computation
for the spectral functions of the vector and axial-vector mesons at finite temperature.
It is seen that
the peaks of the spectral functions
at low temperature will appear like a spike,
the locations of the peak correspond to the groundstate mesons and resonances,
which are found to match with the ones at zero temperature.
We will show how the peaks dissolve
as temperature increasing,
which then indicates the chiral symmetry restoration.
In Sec.\ref{Chap:melt},
we will qualitatively investigate the dissolution of the spectral functions
by fitting to numerical results
in terms of the Breit-Wigner form.
The finite momentum effect in the spectral function is discussed in Sec.\ref{Chap:MD_FME}.
%
%\textcolor{red}{In Sec.\ref{Chap:Const},
At the end of the section, we make remarks on the options in parameters and profiles
for the high order scalar interactions as well as the boundary behavior of the scalar VEV.
%}
%
Our conclusions and remarks are given in Sec.\ref{Chap:Sum}.
Some formula used in our computation of spectral function are derived and presented in\ref{App:HSF}.

\section{Predictive AdS/QCD Model at Finite Temperature}
\label{Chap:Model}

In this section, we will describe the extension of the predictive AdS/QCD model \cite{pADSQCD1,pADSQCD2} to a thermodynamic model by simply considering the black hole metric \cite{Colangelo:2009ra,Misumi}. It then turns out that if taking the previous bulk vacuum expectation value (bVEV) given in \cite{pADSQCD1,pADSQCD2}, the resulting dilaton background will diverge on the horizon at finite temperature due to the modified bulk geometry in the IR-region and the quartic term in the scalar potential, thus a modified bVEV has to be introduced to avoid such a divergence in order to obtain a well-defined spectral function at finite temperature in the predictive soft-wall AdS/QCD model.

Let us begin with considering the following black hole space-time as a bulk gravity:
\begin{equation}\label{metric_dads_01}
ds^2 = a^2(z) \left( f(z) dt^2 - \sum_{i=1}^3 dx^2_i - \frac{dz^2}{f(z)} \right),
\end{equation}
with the IR improved metric and the black hole metric
\begin{eqnarray}\label{f(z)}
a^2(z) = 1/z^2 + \mu_g^2
\quad {\rm and} \quad f(z) &=& 1-(z/z_0)^4.
\end{eqnarray}
Here $z$ has the relation with a black hole radial coordinate $r$ via $z \equiv 1/r$,
it takes from $z_0 \equiv 1/r_0$ (inverse horizon) to $0$ (boundary).
(In Sec.\ref{Chap:VM}, we will see that it is more convenient to use the variable $u \equiv z/z_0$.)
As the boundary theories are independent of the AdS radius, we will set it as unity through this paper.
The Hawking temperature of the black hole is given as $T = 1/(z_0\pi)$,
which corresponds to the temperature in boundary theories. The mass scalae $\mu_g$ characterizes QCD confinement and is fixed
by the low energy physics of hadrons\cite{pADSQCD1,pADSQCD2}.

%\textcolor{red}{
We take the black hole space-time as the bulk geometry,
despite what the boundary theory is in confinement phase, as in the field theory
the mesons are considered as the bound states of strong interaction. In our treatment, the bulk geometry of
the black hole space-time given in Eq. (\ref{metric_dads_01}) is taken as the background.
The dilaton is also taken as the background field, but it is obtained in the bulk geometry of the black hole space-time
from solving the equation of motion for the scalar field with the required boundary conditions, the details can be seen below.
Thus, our system can be a solution of equation of motion in the end.
%---
(While it is not known whether it is stable or metastable.
One way for the confirmation of that is via
the evaluation of the specific heat like ref.\cite{Miranda:2009uw}.
)
%---
This is the different point from other studies\cite{Misumi,Colangelo:2009ra,Miranda:2009uw},
where both of the dilaton and the black hole space-time are taken as the backgrounds for confinement phase.
%---
%}

%\textcolor{red}{
%The holographic model dual to confinement phase of the realistic QCD would be at finite temperature, if it were.
%It would be some AdS-type or not too different from AdS.
%Our model could be considered as some effective model for that full model.
%}

On the background geometry Eq.(\ref{metric_dads_01}), the considered field ingredients include the dilaton background $\Phi$,
the bulk scalar field $X$ and gauge fields $A_L$ and $A_R$ of $SU_L(2)\times SU_R(2)$ gauge symmetry.
Here $X = X^a t^a$ and $A_{L,R} = A^a_{L,R} t^a$ with $t^a$ ($a=1,2,3$) the $SU(2)$ Lie algebra are
as the dual operators in the boundary theories. The model is given as follows on the background Eq.(\ref{metric_dads_01}),
\begin{eqnarray}\label{lag}
S &=& S_\Phi + S_{\rm gauge},
\end{eqnarray}
where
\begin{eqnarray}
S_\Phi
&=&
\int d^5x \, \sqrt{g}e^{-\Phi(z)}\,{\rm{Tr}} \Big( |D_M X|^2 -V(X) \Big),
\label{lagP}\\
S_{\rm gauge}
&=& - \frac{1}{4g_5^2}
\int d^5x \, \sqrt{g} e^{- \Phi(z)}  \Tr
\bigl( F_{L,MN}F_{L}{}^{MN} + F_{R,MN}F_{R}{}^{MN}  \bigr),
\label{lagG}
\end{eqnarray}
with
\begin{eqnarray}
D_M X &=& \partial_M X + i\left(A_{L,M} X - X A_{R,M}\right),\\
    V(X) &=& m_X^2 |X|^2 + \frac{\lambda}{4} |X|^4.  \label{qtt}
\end{eqnarray}
with $m_X^2=-3$. $g_5^2 \equiv 12\pi^2L/N_c $ with $N_c =3$ is a gauge coupling given in \cite{Karch:2006pv}.
We use capital Latin and Greek character for the five dimensional and the four dimensional coordinates, respectively,
$M=x_0,x_1,x_2,x_3,z$ and $\mu=x_0,x_1,x_2,x_3$.
$V(X)$ is the potential of the bulk scalar field. $\lambda$ is a coefficient for the quartic term.
Here we will consider the case with or without the quartic term, i.e., $\lambda =0$ and $\lambda \neq 0$. For $\lambda \neq 0$, its value at zero temperature was fitted from the meson mass spectra and found to be $\lambda  = 9$ \cite{pADSQCD1,pADSQCD2}.

Let us now discuss the bulk vacuum expectation value (bVEV) of the scalar field and the dilaton background at finite temperature.
Taking the bVEV to be a function of the fifth dimension $v(z)$
\begin{eqnarray}\label{X(z)}
X(z) = \frac{1}{2}\,v(z)~\mathbf{1}_2,
\end{eqnarray}
where $\mathbf{1}_2$ denotes $2 \times 2$ unit matrix.
$v(z)$ at zero temperature is given by Table.\ref{Tbl01}, \ref{Tbl02wo} and \ref{Tbl02w}.
Here we refer it as zero temperature part and denote it as $v_0(z)$ below.

%
%%%%%%%%%%%%%%%%%%%%%%%%%%%%%%%%%%%%%%%%%%
%
\begin{table}[!h]
\begin{center} \begin{tabular}{ c l l l l }
\hline\hline
Model & $\qquad\qquad$ $v_0(z)$  & \multicolumn{3}{c}{Parameters}   \\
\hline
Ia  & $ z(A+B z^2)(1+C z^2)^{-1} $  & $A = m_q \zeta$,& \quad $B=\sigma/{\zeta}+m_q\zeta C$,            & \quad  $C=B / (\mu_d \gamma) $\\
Ib  & $ z(A+B z^2)(1+C z^2)^{-5/4}$ & $A = m_q \zeta$,& \quad $B=\sigma/{\zeta}+\frac{5}{4}m_q\zeta C$, & \quad  $C=(B^2/ (\mu_d \gamma^2))^{2/5}$ \\
IIa & $ z(A+B z^2)(1+C z^4)^{-1/2}$ & $A = m_q \zeta$,& \quad $B=\sigma/{\zeta}$,                       & \quad  $C=(B/( \mu_d\gamma))^2$ \\
IIb & $ z(A+B z^2)(1+C z^4)^{-5/8}$ & $A = m_q \zeta$,& \quad $B=\sigma/{\zeta}$,                       & \quad  $C=(B^2/(\mu_d\gamma^2))^{4/5}$ \\
\hline\hline
\end{tabular}
\caption{
These are taken from \cite{pADSQCD1}.
Numerical values are given in Table.\ref{Tbl02wo} and \ref{Tbl02w}.
}
\label{Tbl01}
\end{center}
\end{table}
%
%%%%%%%%%%%%%%%%%%%%%%%%%%%%%%%%%%%%%%%%%%
%
\begin{table}[!h]
\begin{center}
\begin{tabular}{ccccc}
\hline\hline
Model & $m_q$ (MeV) & $\sigma^{\frac{1}{3}}$ (MeV) & $\mu_g$ (MeV)  & $\gamma$ \\
\hline
Ia  & 4.16 & 275 & $(2\mu_d^2) / 3$ & 0.178\\
Ib  & 4.64 & 265 & $\mu_d^2 / 3$    & 0.136\\
IIa & 4.44 & 265 & $(2\mu_d^2) / 3$ & 0.153\\
IIb & 4.07 & 272 & $\mu_d^2 / 3$    & 0.112\\
\hline\hline
\end{tabular}
\caption{
The numerical values in the case without the quartic term ($\lambda_0=0$) \cite{pADSQCD1}.
$\zeta=\sqrt{3}/(2\pi)$ and $\mu_d=445 $ MeV for all the models.}
\label{Tbl02wo}
\end{center}
\end{table}
%
%%%%%%%%%%%%%%%%%%%%%%%%%%%%%%%%%%%%%%%%%%
%
\begin{table}[!h]
\begin{center}
\begin{tabular}{ccccccc}
\hline\hline
Model & $m_q$ (MeV) & $\sigma^{\frac{1}{3}}$ (MeV) &  $\mu_d$ (MeV) & $\mu_g^2$ (MeV$^2$) & $\gamma$   \\
\hline
IIa   & $6.95$ & $228$ &  412 & $(2\mu_d^2) / 3$ & $0.30$ \\
IIb   & $6.79$ & $229$ & 548 & $\mu_d^2 / 3$ & $0.20$ \\
\hline\hline
\end{tabular} 
\caption{
The numerical values in the case with the quartic term ($\lambda_0=9$) \cite{pADSQCD1}.
$\zeta=\sqrt{3}/(2\pi)$ for all the models.
}
\label{Tbl02w}
\end{center}
\end{table}
%
%%%%%%%%%%%%%%%%%%%%%%%%%%%%%%%%%%%%%%%%%%
%

At finite temperature characterized by the black hole metric, the dilaton field $\Phi(z)$ has to be consistently determined from solving the equation of motion for any known bVEV $v(z)$
\begin{eqnarray}\label{eomPhi2}
\Phi'(z) =
\frac{3a'(z)}{a(z)}
+
\frac{\big(f(z)v'(z)\big)'}{f(z )v'(z)}
%- \frac{3}{z} +\frac{3\Omega'(z)}{\Omega(z)}
-\frac{a^2(z)}{f(z)v'(z)}
\left(
m_X^2v(z)+\frac{\lambda}{2}v^3(z)
\right),
\end{eqnarray}
where the prime ($'$)  means a $z$-derivative. It is seen that there are terms with $f(z)$ in the denominator in eq.(\ref{eomPhi2}),
which leads to a divergent solution for the dilaton field $\Phi(z)$ on the horizon ($z \to 1$) at finite temperature
if taking the bVEV given in \cite{pADSQCD1} for zero temperature case. While the divergent solution of $\Phi(z)$ will bring
the difficulty in taking the in-falling boundary condition for the spectral functions. Note that the divergence appears only on the horizon at finite temperature due to the black hole metric $f(z)$. At zero temperature $f(z)=1$.

To avoid such a divergence in the solution for the dilaton field $\Phi(z)$, we may modify the bVEV $v(z)$ to be
\begin{eqnarray}\label{vassume}
v(z) = v_1(z) \ln f(z) + v_0(z),
\end{eqnarray}
where $v_0(z)$ is taken to be the part at zero temperature and given in \cite{pADSQCD1}. The part $v_1(z)\ln f(z)$ vanishes at zero temperature and
is introduced to obtain a finite solution for the dilaton field $\Phi(z)$ by a proper $v_1(z)$.  To fix $v_1(z)$, we substitute the above modified bVEV into eq.(\ref{eomPhi2}), which gives
\begin{eqnarray} \label{dphi}
\Phi'(z) &=&
\frac{\ln f(z)}{v_1(z) f'(z)}
\left\{
- a^2(z) v_1(z)
\left(
m_X^2 + \frac{\lambda}{2}v^2(z)
\right)
+ f'(z) v_1'(z)
\right\} + \cdots, \nonumber \\
& = & \frac{\ln f(z)}{v_1(z) f'(z)}
\left\{
{\cal F}(z)
-  \frac{\lambda}{2} a^2(z) v_1(z) v^2(z)
\right\} + \cdots.
\end{eqnarray}
here we write only the divergent term proportional to $\ln f(z)$,
and ``$\cdots$'' denotes the finite terms at the horizon limit: $\displaystyle z \to 1$.
Where we have defined the function ${\cal F}(z)$ as
\begin{eqnarray}
{\cal F}(z) \equiv - m_X^2 a^2(z) v_1(z) + f'(z) v_1'(z)
\end{eqnarray}
It is seen that for the case $\lambda=0$ the solution for  $\Phi(z)$ becomes finite once ${\cal F}(z)=0$.
It is not difficulty to find from the condition ${\cal F}(z)=0$ that $v_1(z)$ should take the following form
\begin{eqnarray} \label{v1}
v_1(z) = c_v \exp\left[  \frac{m_X^2}{8(\pi T)^4z^4} \left( \frac{1}{2} + \mu_g^2 z^2 \right)\right],
\end{eqnarray}
with $c_v$ an integral constant. The above result holds for all the models discussed in \cite{pADSQCD1}.
We will make a detailed discussion below for the fixing of the constant $c_v$.

For the case $\lambda\neq 0$, there is an additional divergent term in Eq.(\ref{dphi}). When taking Eq.(\ref{v1}), the divergent term is given as
\begin{eqnarray} \label{dphi2}
\Phi'(z) &=& -
\frac{\ln f(z)}{f'(z)}
\frac{\lambda}{2}a^2(z)v^2(z)
+ \cdots.
\end{eqnarray} 
its divergent order is seen to be $(\ln f(z))^3$. As a simple prescription, we may take $\lambda$ as the following form
\begin{eqnarray}\label{asu2}
\lambda = \frac{\lambda_0}{1+ c_\lambda (\ln f(z))^p} \qquad {\rm with} \qquad p \ge 3,
\end{eqnarray}
with $c_\lambda$ a constant. Thus the resulting solution for $\Phi(z)$ becomes finite as long as $p \ge 3$. For simplicity,
we take $p=3$ in the present consideration.

\begin{figure}[!h]
\begin{center}
\includegraphics[width=64mm,clip]{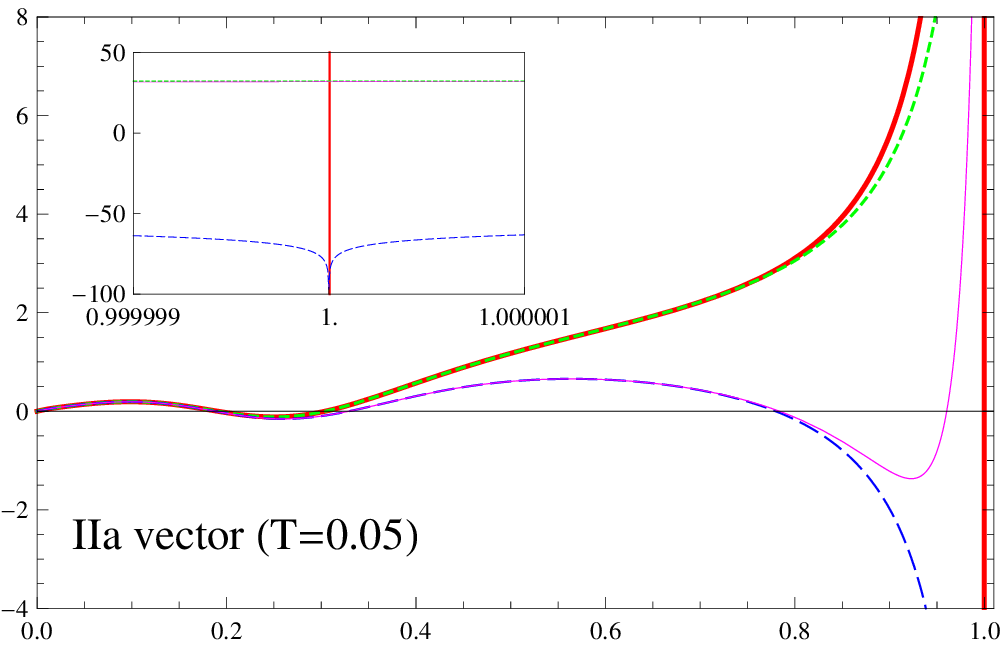}
\includegraphics[width=64mm,clip]{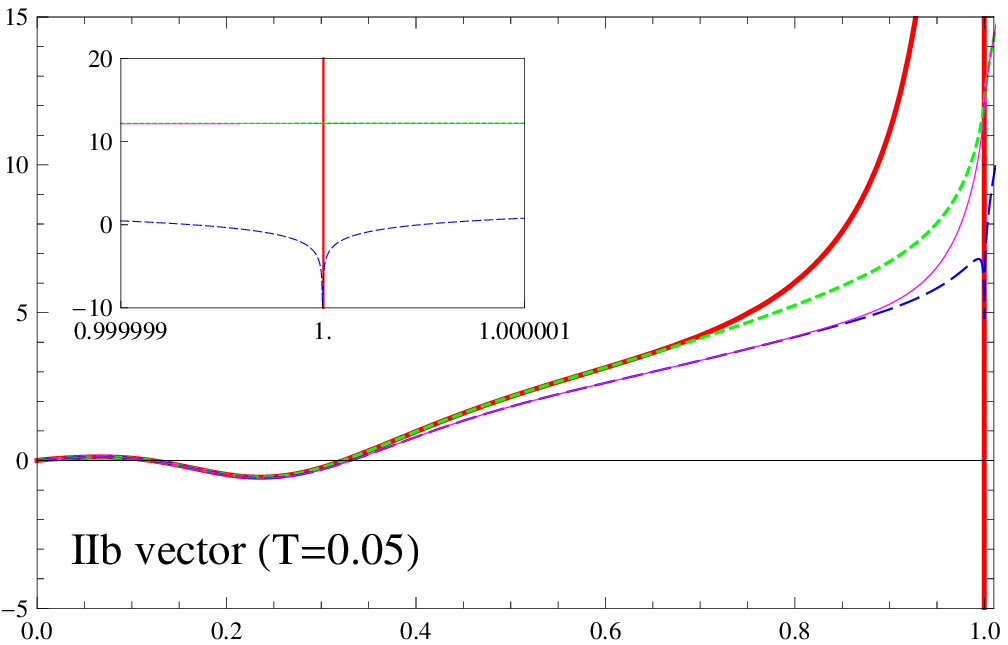}
\includegraphics[width=64mm,clip]{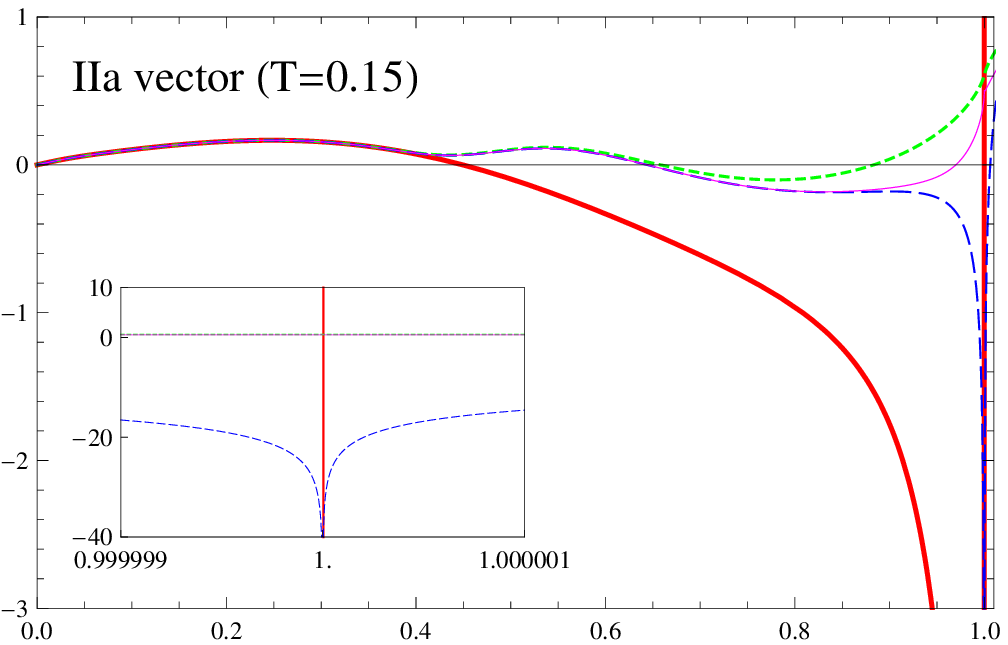}
\includegraphics[width=64mm,clip]{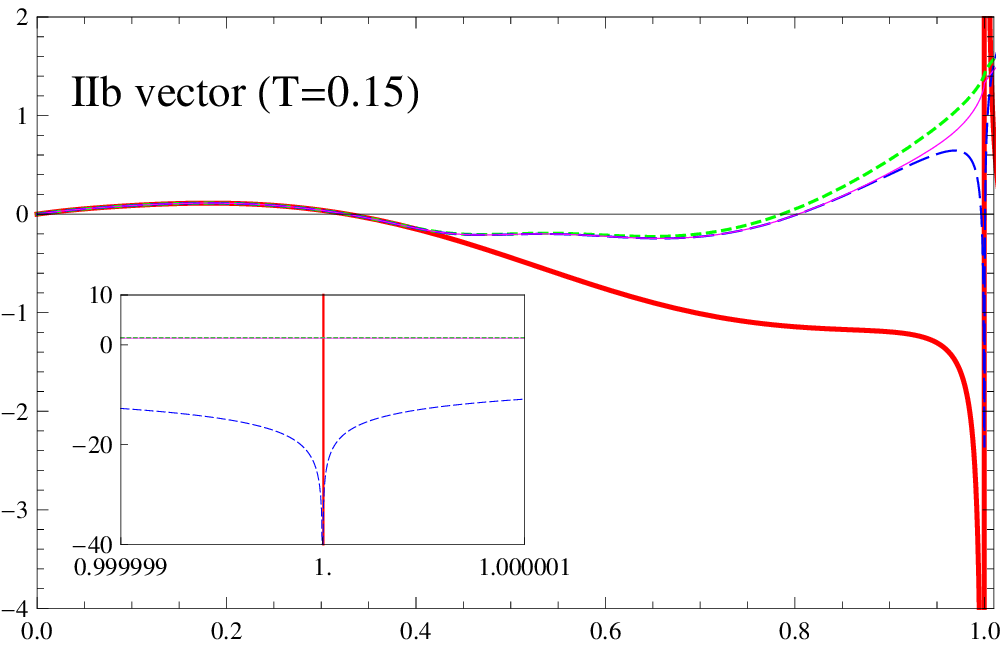}
\end{center}
\caption{ The x- and y-axis mean $u=z/z_0$ and $\Phi'(u)$. $\Phi'(u)$ is shown to be well-defined in the bulk entirely
with the bVEV in Eq.(\ref{vassume}) and the quartic coupling in Eq.(\ref{asu2}).  $T=0.05$ and $0.15$ GeV correspond to low and high temperatures in this paper. The correspondence between lines and parameters is given in Table.\ref{Tblclr}.
Where the small windows give the magnified vicinity of the horizon, the red line (or thin solid line) and the blue line (or dashed line) show the divergence with behaviors as $1/u^2$ and cubic logarithmic as expected from Eqs.(\ref{eomPhi2}) and (\ref{dphi2}) for $c_v=0$ and $c_\lambda=0$. When $c_v$ (green or dotted line) and $c_\lambda$ (magenta or thick solid line) are turned on, the divergences disappear. In the small windows,
green line (dotted line) and magenta line (thick solid line) are overlapping each other.
}
\label{Fig_regudPhi}
\end{figure}

With the modified bVEV given in Eq.(\ref{vassume}) and the modified coefficient of the quartic term given in Eq. (\ref{asu2}),
we then arrive at a nonsingular solution for $\Phi'(z)$ as shown in Fig.\ref{Fig_regudPhi}.
\newline

%
%%%%%%%%%%%%%%%%%%%%%%%%%%%%%%%%%%%%%%%%%%
%
%
%%%%%%%%%%%%%%%%%%%%%%%%%%%%%%%%%%%%%%%%%%
%
\begin{table}[!h]
\begin{center} \begin{tabular}{ r|ccc }
\hline\hline
    line \qquad \qquad & $c_v$ & $c_\lambda$ & $\lambda_0$ \\
\hline
    red (thin solid)  & 0    & 0    & 0 \\
  green (dotted)      & -0.1 & 0    & 0 \\
   blue (dashed)      & -0.1 & 0    & 100 \\
magenta (thick solid) & -0.1 & -0.1 & 100 \\
\hline\hline
\end{tabular}
\caption{ The correspondence between the lines and the parameters used in Fig.\ref{Fig_regudPhi}.
Other parameters are given in Table.\ref{Tbl02w}. The red line (thin solid line) and the blue line (dashed line) diverge
as expected from Eqs.(\ref{eomPhi2}) and (\ref{dphi2}). Where the green line (dotted line) and the magenta line (thick solid line) correspond to the modified bVEV given in Eq.(\ref{vassume}) and the modified coefficient given in Eq.(\ref{asu2}).
}
\label{Tblclr}
\end{center}
\end{table}
%
%%%%%%%%%%%%%%%%%%%%%%%%%%%%%%%%%%%%%%%%%%
%
%\textcolor{red}{
Now let us further discuss the parameters $c_v$ and $c_\lambda$.
As they cannot be determined theoretically in the present considerations,
we are going to fix them from the phenomenological studies.
%-----
Their numerical values will be given in the next section, here we shall first discuss the general considerations for how to fix them.
%}

%\textcolor{red}{

It is seen that the parameter $c_v$ is introduced as the coefficient of the term  $v_1(z)$ given in Eq.(\ref{v1})
which is associated with the part of the VEV $v_1(z) \ln f(z)$ in Eq.(\ref{vassume}). This part of the VEV plays
the role of a regulator, otherwise the solution of dilaton will be divergent. Similarly, the parameter $c_{\lambda}$ is introduced
in Eq. (\ref{asu2}) to modify the zero-temperature coupling of the quartic term of the bulk scalar in order to yield a convergent
solution for the dilaton.
%-----
It is noticed that as long as the parameters $c_v$ and  $c_{\lambda}$ are non-zero, even if no matter how small they are,
the divergence at the horizon can be removed.
%-----
On the other hand, as two parameters $c_v$ and  $c_{\lambda}$ become smaller and smaller,
the differences from the zero-temperature part
given as $v_0(z)$ in Eq.(\ref{vassume})
also become smaller and smaller.
%-----
Thus we will assign values to $c_v$ and  $c_{\lambda}$ as small as possible.
%By doing so, we can limit the role of these just only as a regulator.
%}

%\textcolor{red}{
For simplicity, we will treat $c_v$ and $c_\lambda$ equally as $c_v=c_\lambda$.
In general, they can be a function of temperature, but we are going to treat them as a constant also for simplicity.
%}

\section{Spectral function for vector and axial-vector mesons}
\label{Chap:VM}

To begin with, we combine the gauge fields
into vector field $V^a_M$ and axial-vector field $W^a_M$ as
\begin{eqnarray}\label{def_VW}
V^a_M \equiv \frac{1}{2}(A^a_{L,M}+A^a_{R,M})
\quad {\rm and} \quad
W^a_M \equiv \frac{1}{2}(A^a_{L,M}-A^a_{R,M}),
\end{eqnarray}
%---
where $M =x_0,x_1,x_2,x_3,z$ as mentioned below eq.(\ref{lag}).
%---
The dual current operators for $V^a_M$ and $W^a_M$
are flavor vector current $\bar{q}\gamma^\mu t^a q$ and
flavor axial-vector current $\bar{q}\gamma^5 \gamma^\mu t^a q$, respectively.
%---
For convenience, we will change the radial coordinate $z(\equiv 1/r)$ to $u$ as
\begin{eqnarray}\label{defu}
u \equiv z/z_0 = (\pi T)z.
\end{eqnarray}

The equations of motion for the vector field $V^a_x$ and the axial-vector field $W^a_x$
in the spatial component $x=x_1,x_2,x_3$ are given as
\begin{eqnarray}
\label{eq:eomV1}
\textrm{V} \, &:& \,
0 =
\partial_u \Bigl[ e^{-\Phi(u)} \sqrt{g}\, g^{xx} g^{uu}\, \p_u V^a_x(x,u) \Bigr]
+
\Bigl[ e^{-\Phi(u)} \sqrt{g}\, g^{xx} \partial_\mu \partial^\mu V^a_x(x,u) \Bigr],\\
%=====
\label{eq:eomAV1}
\textrm{AV} \, &:& \,
0 =
\p_u\Bigl[ e^{-\Phi(u)} \sqrt{g}\, g^{xx} g^{uu}\, \p_u W^a_x(x,u) \Bigr]
+
\Bigl[ e^{-\Phi(u)} \sqrt{g}\, g^{xx} \p \p^\mu W^a_x(x,u) \Bigr] \nonumber
\\ && \qquad\qquad
+~4  g_5^2 g_{uu} ( X^0(u) )^2 W^a_x(x,u),
\end{eqnarray}
where ``V'' and ``AV'' denote the vector and axial-vector, respectively.
%---
As in ref.\cite{Misumi},
we adopt the axial gauge condition
for the radial direction as $A_{L,z}=A_{R,z}=0$
and the Landau gauge for the 4d space-time
as $\partial^\mu A_{L,\mu}=\partial^\mu A_{R,\mu}=0$
to get rid of unphysical polarization.
%---

Performing Fourier transformation for the four-dimensional coordinates as
$\displaystyle V^a_x(x,u) = \int d^4 p \, e^{i p \cdot x} \widetilde{V}^a_x(p,u)$
and
$\displaystyle W^a_x(x,u) = \int d^4 p \, e^{i p \cdot x} \widetilde{W}^a_x(p,u)$
with $p_\mu=(\omega,q_1,q_2,q_3)$ and the momentum $\vec{q}$ taken as one direction,
we have
\begin{eqnarray}
\label{eq:eomV2}
&&  \widetilde{V}^a_x{}''(p,u)
+  F_1(u) \widetilde{V}^a_x{}'(p,u)
+ \frac{1}{( \pi T)^2} F_0(u) \widetilde{V}^a_x(p,u)=0,\\
%=====
\label{eq:eomAV2}
&&\widetilde{W}^a_x{}''(p,u)
+ F_1(u) \widetilde{W}^a_x{}'(p,u)
+ \frac{1}{( \pi T)^2} F_0(u) \widetilde{W}^a_x(p,u) + ~g_5^2 \frac{ v^2(u)}{u^2 f(u)} \widetilde{W}^a_x(p,u) =0.
\end{eqnarray}
with
\begin{eqnarray}
F_1(u) = \frac{a'(u)}{a(u)} + \frac{f'(u)}{f(u)} - \Phi'(u), \qquad F_0(u)  =  \frac{\omega^2}{f^2(u)} - \frac{q^2}{f(u)}
\end{eqnarray}
where the prime ($'$) means a derivative with regard to $u$. It is seen that there is no difference
in the solutions $\widetilde{V}^a_x(p,u)$ and $\widetilde{W}^a_x(p,u)$ coming from the index for $SU(2)$ algebra $a$, so we will not attach the index $a$ to the spectral function below.

With the above analysis, we are now in the position to make numerical computation for
the retarded two-point Green function $G^a(\omega,q)$
for the operators of flavor vector current or flavor axial-vector current.
Then we can get the spectral function as
\begin{equation} \label{eq:spectralv}
\rho(\omega,q)=-\frac{1}{\pi} \mathrm{Im} \, G^a(\omega,q) \, \theta(\omega^2-q^2).
\end{equation}
Here the index $a$ will not be attached to $\rho(\omega,q)$
for the reason mentioned above. where $\theta(\omega^2-q^2)$ is a step function
introduced to keep the squared energy being positive in the boundary theory
as a general requirement likewise ref.\cite{Kovst}.
As the key point in spectral functions is
the location and the width of the peak,
where the width and the location of the peak are defined in our present study by the Breit-Wigner form given in eq.(\ref{fitfunc}),
we ignore the overall factor $-1/\pi$ in our actual analysis for simplicity.

The solutions $\widetilde{V}^a_x(q,u)$ and $\widetilde{W}^a_x(q,u)$
described by the equations of motion (\ref{eq:eomV2}) and (\ref{eq:eomAV2})
can be generally given by two linear independent solutions as
\begin{eqnarray} \label{eq:Asy44}
\quad \widetilde{V}^a_x(p,u) = A^a(\omega,q) \Phi^a_0(\omega,q,u) + B^a(\omega,q) \Phi^a_1(\omega,q,u).
\end{eqnarray}
As the axial-vector sector is the same as the vector sector, we will skip its description below.
As described in appendix, the ratio $B(\omega,q)/A(\omega,q)$
corresponds to the retarded two-point Green function,
which can be written as
\begin{eqnarray} \label{BoA}
G^a(\omega,q)
=
C \frac{B^a(\omega,q)}{A^a(\omega,q)} + \cdots
= \frac{C}{\Phi^a_1(\omega,q,u)} \frac{\widetilde{V}^a_x(\omega,q,u)}{\widetilde{V}^a_x(\omega,q,0)} + \cdots.
\end{eqnarray}
Here we have assumed that
$\Phi^a_0$ becomes some constant
and $\Phi^a_1$ vanishes on the boundary ($u \to 0$).
The part ``$\cdots$'' represents irrelevant part in our present consideration.
$C$ represents an overall constant fixed from QCD.

Our approach to get the Green function (\ref{BoA}) is the shooting method.
For that, we need the asymptotic behaviors for the boundary conditions.
It turns out that the asymptotic behaviors are the same for all the models though the dilaton takes different form for the different models when solving the equations of motion (\ref{eq:eomV2}) and (\ref{eq:eomAV2}). Thus, the following description is common for all models.
From Eqs.(\ref{eq:eomV2}) and (\ref{eq:eomAV2}), we can obtain the solutions for $\Phi^a_0$ and $\Phi^a_1$ at the vicinity of the boundary
in the limit: $z \to \epsilon$,  with $\epsilon$ an infinitesimal number. The solutions are composed of two liner-independent functions by the first-kind Bessel function $J_1$ and second-kind Bessel function $Y_1$, we assign them as follows
\begin{eqnarray} \label{eq:Asy45}
\textrm{V} \, &:& \,
\Phi^a_0(\omega,q,u=\epsilon) = \epsilon~Y_1 ( x_{\rm V} )
\quad {\rm and} \quad
\Phi^a_1(\omega,q,u=\epsilon) = \epsilon~J_1 ( x_{\rm V} ), \\
\textrm{AV}  \, &:& \,
\Phi^a_0(\omega,q,u=\epsilon) = \epsilon~Y_1 ( x_{\rm AV} )
\quad {\rm and} \quad
\Phi^a_1(\omega,q,u=\epsilon) = \epsilon~J_1 ( x_{\rm AV} ),
\end{eqnarray}
with
$\displaystyle
x_{\rm V} \equiv \frac{\sqrt{\omega^2-q^2}}{\pi T} \epsilon
$ and
$
x_{\rm AV} \equiv \frac{\sqrt{\omega^2-q^2 -g_5^2A^2}}{\pi T} \epsilon
$.
On the other hand, solving Eqs.(\ref{eq:eomV2}) and (\ref{eq:eomAV2}) around the horizon in the limit: $u \to 1-\epsilon$,
we find that the solutions for the vector and axial-vector mesons
are the same with the so called in-going and out-going solution.
As usual, we take the in-going solution as the boundary condition:
\begin{equation} \label{eq:Asy46}
\widetilde{V}^a_x(\omega,q,u=1-\epsilon) = \widetilde{W}^a_x(\omega,q,u=1-\epsilon) = \epsilon^{- i \frac{\omega}{4 \pi T}}.
\end{equation}

We now make numerical computation with fixing the values for $c_v$ and $c_\lambda$ and considering the temperature effect.
In principle, one can assign their values as small as possible, while in the practical numerical computations, it is difficult
to take extremely small values for both $c_v=c_\lambda$ and temperature $T$. It will be seen that the mass spectrum appears sharply like a spike at low temperature, while as the temperature turns on, the spike becomes wider and wider as well as lower and lower when the temperature
gets higher and higher. In order to check such a behavior, we will start with the temperature as low as possible, and turn on temperature gradually.
In the practical computations, the possible lowest temperature is assigned in the following figures for the different cases.
In figuring out the values of $c_\lambda$ and $c_v$, it is seen  that the positive value is excluded, which can be seen from  Eq.(\ref{eomPhi2}) or Eq.(\ref{asu2}), for a positive value, a divergence appears in $\Phi'(u)$ at somewhere in $u$ moving from $0$ to $1$.
Thus, we will consider only the negative values for $c_\lambda$ and $c_v$. As a consequence, the possible small values for $c_\lambda$, $c_v$ are assigned in the practical computations to be as follows :
\begin{equation} \label{v1v}
\begin{split}
\textrm{The case without the quartic term ($\lambda_0=0$)}
\, &: \,
c_v= -10^{-3} \quad \textrm{(constant)}. \\
%---
\textrm{The case with the quartic term ($\lambda_0=9$)}
\, &: \,
c_v=c_\lambda= - 10^{-4} \quad \textrm{(constant)}.
\end{split}
\end{equation}

\begin{figure}[!h]
\begin{center}
\includegraphics[width=64mm,clip]{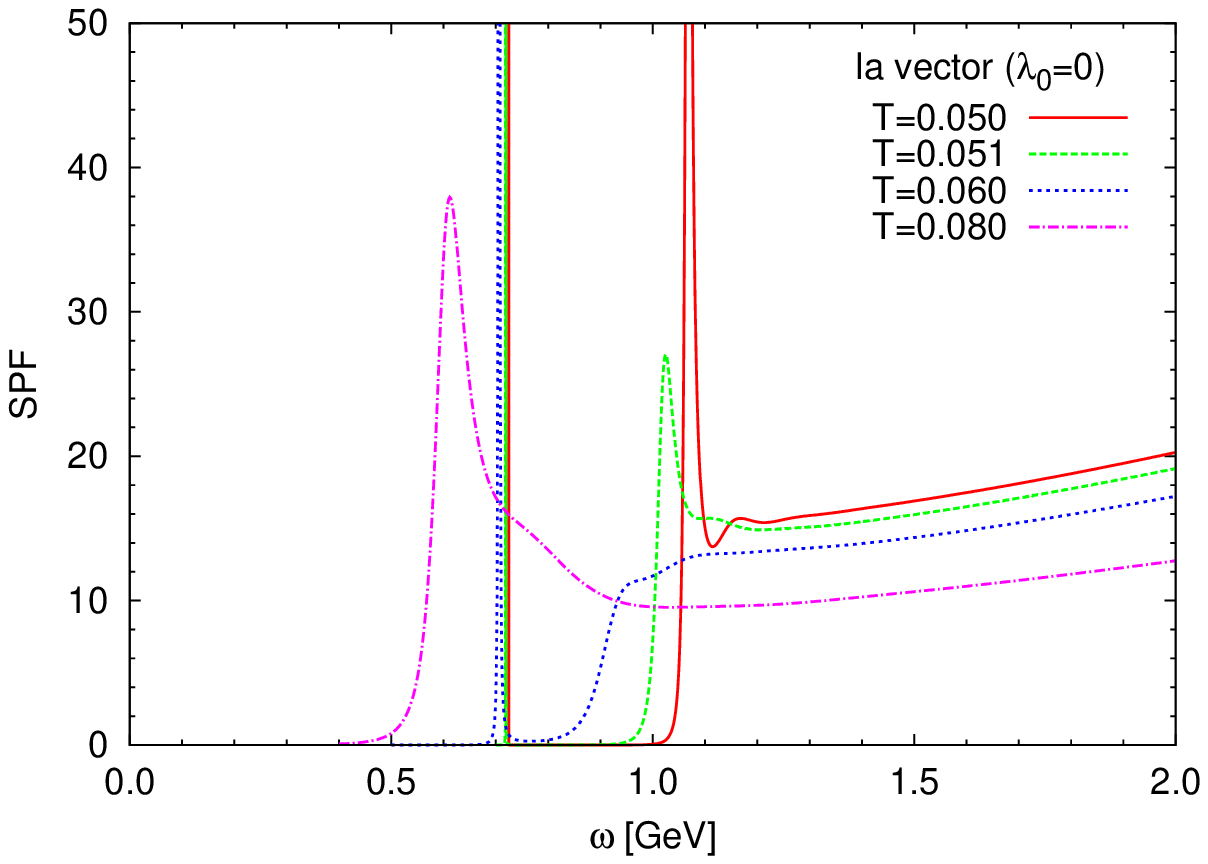}
\includegraphics[width=64mm,clip]{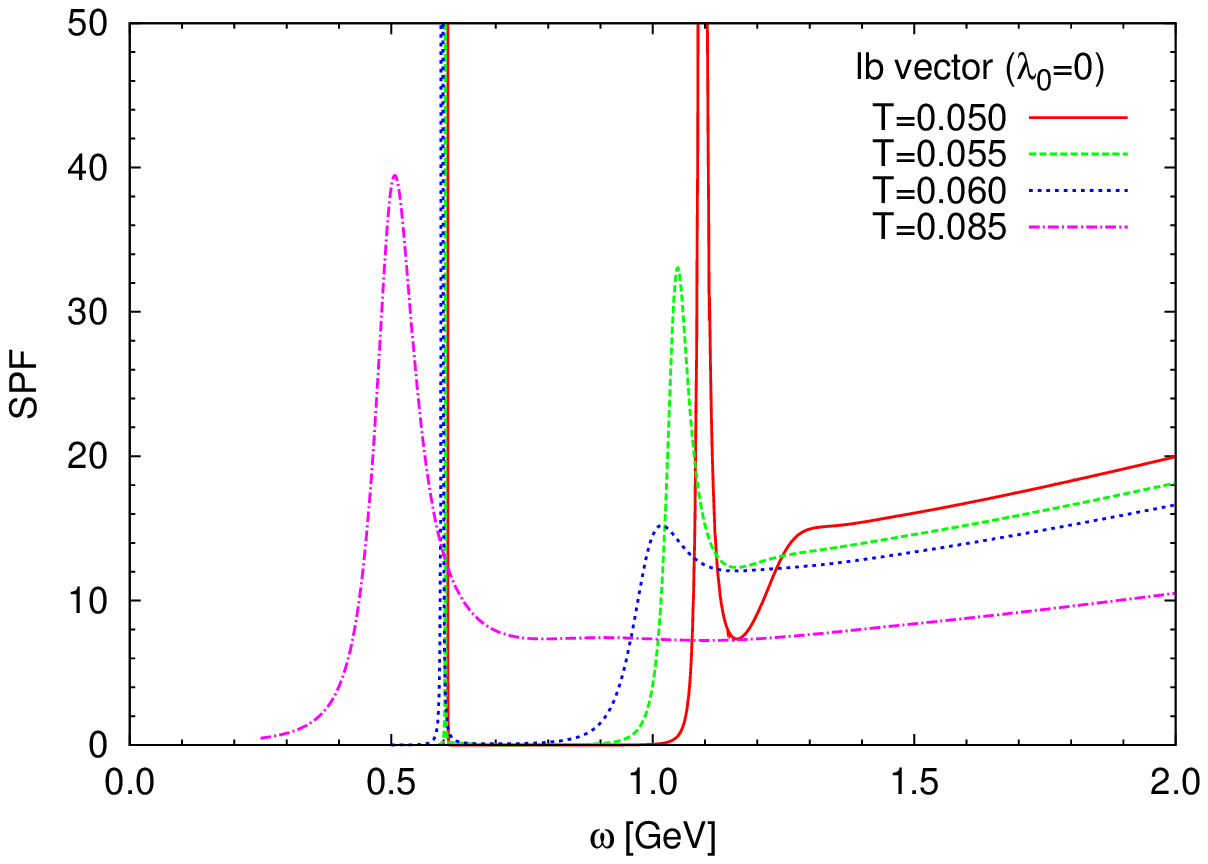}
\includegraphics[width=64mm,clip]{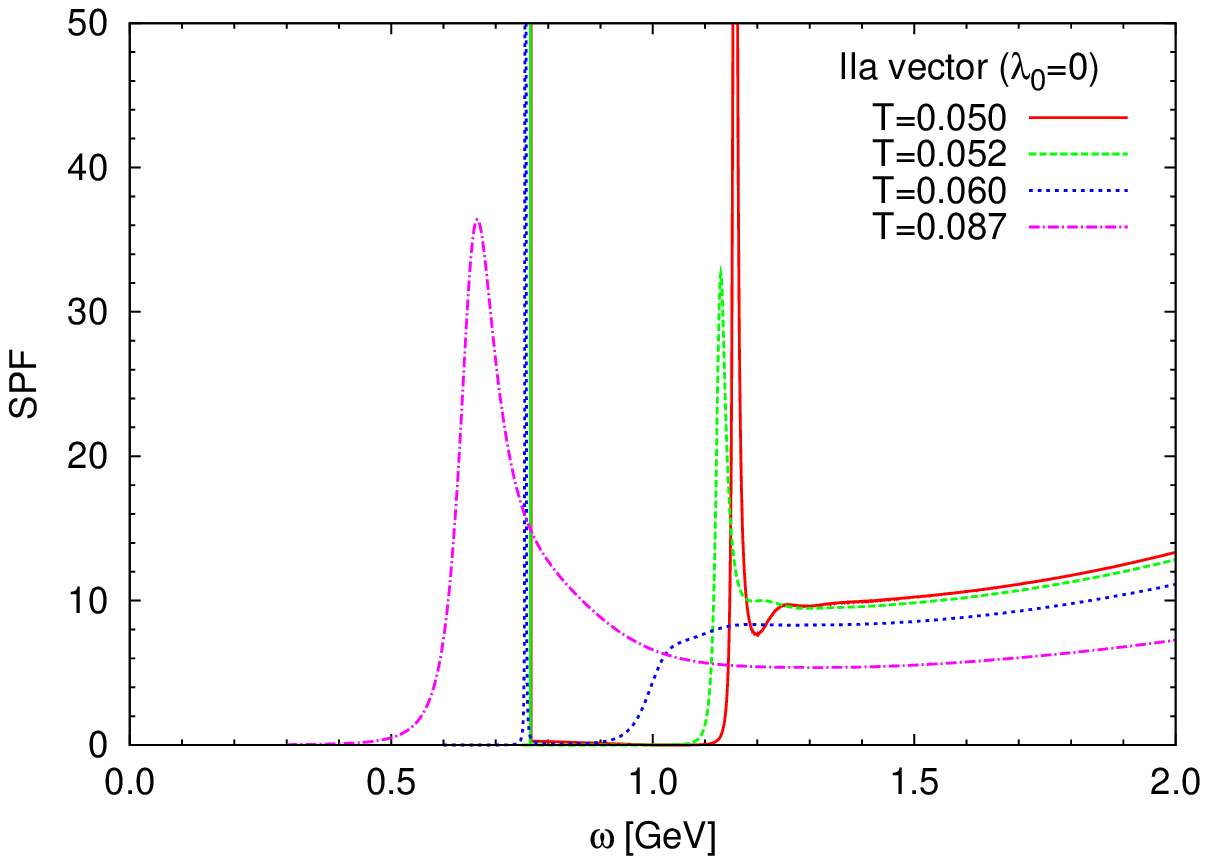}
\includegraphics[width=64mm,clip]{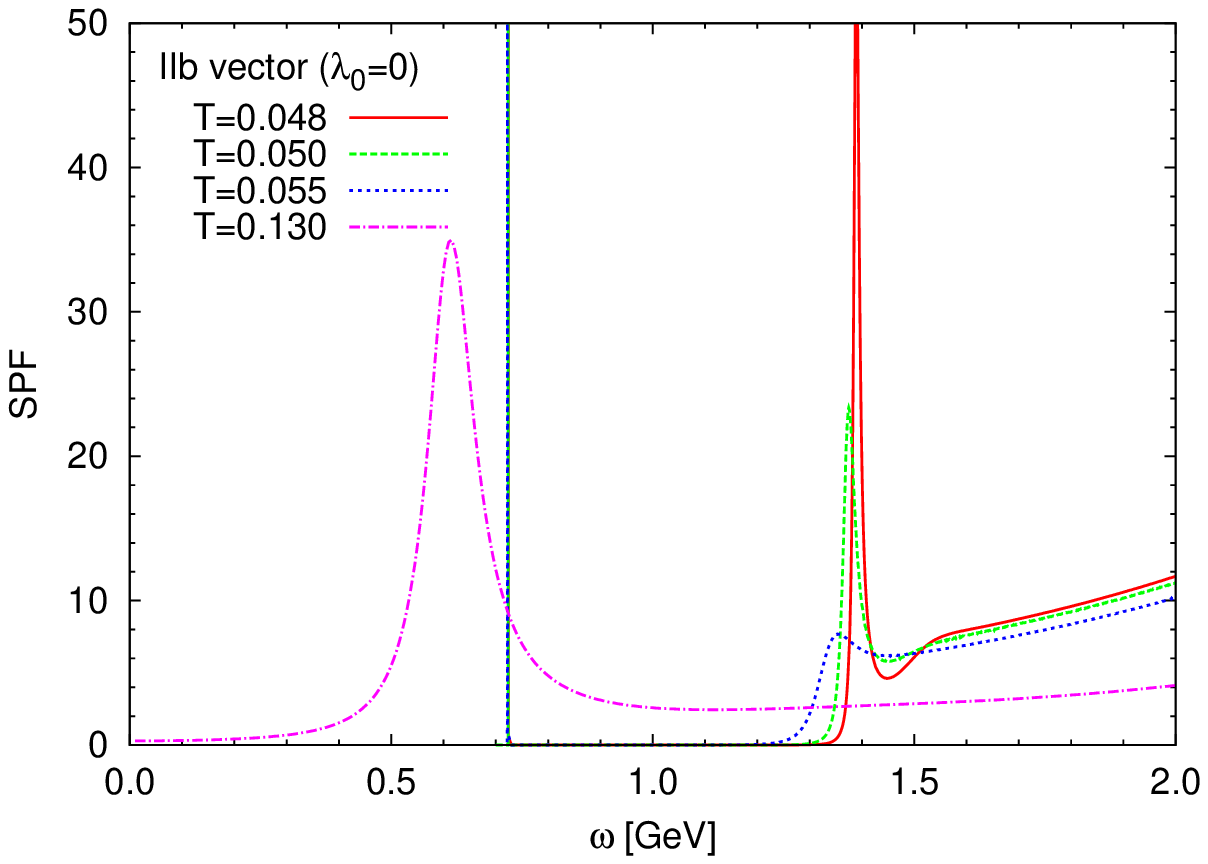}
%-----
\end{center}
\caption{
The results of the spectral function for vector meson
defined in eq.(\ref{eq:spectralv})
without the quartic term ($\lambda_0=0$).
As $q=0$, the positions of the peaks in x-axis can be read as masses of mesons.
%---
The parameters used here are given in eq.(\ref{v1v}) and Table.\ref{Tbl02wo},
and the unit of temperature is in GeV.
The radial coordinate is taken as $u \in [1-10^{-8},~10^{-8}]$ in the numerical computation.
%---
The locations of the spike match with the results shown in Table.\ref{vecm_nqt}.
}\label{Fig_vec_nq}
\end{figure}

\begin{figure}[!h]
\begin{center}
\includegraphics[width=64mm,clip]{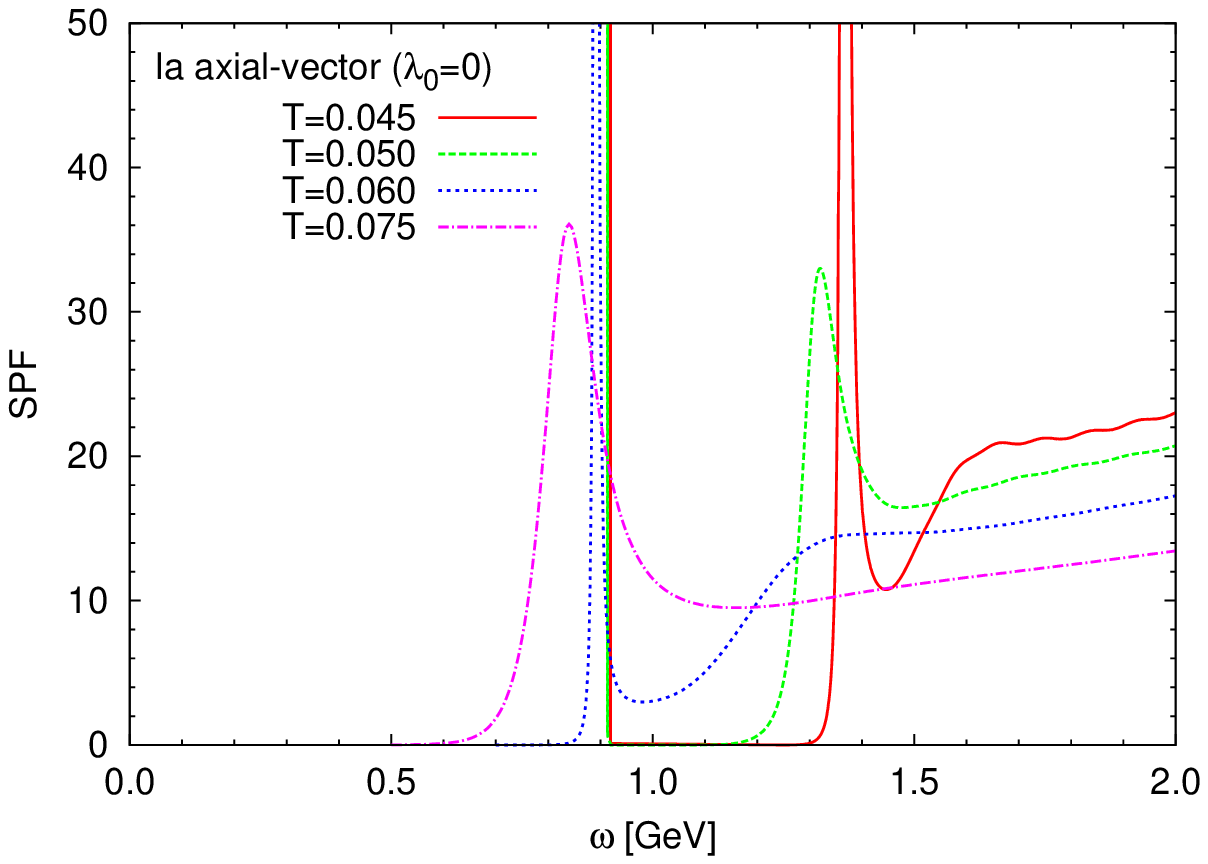}
\includegraphics[width=64mm,clip]{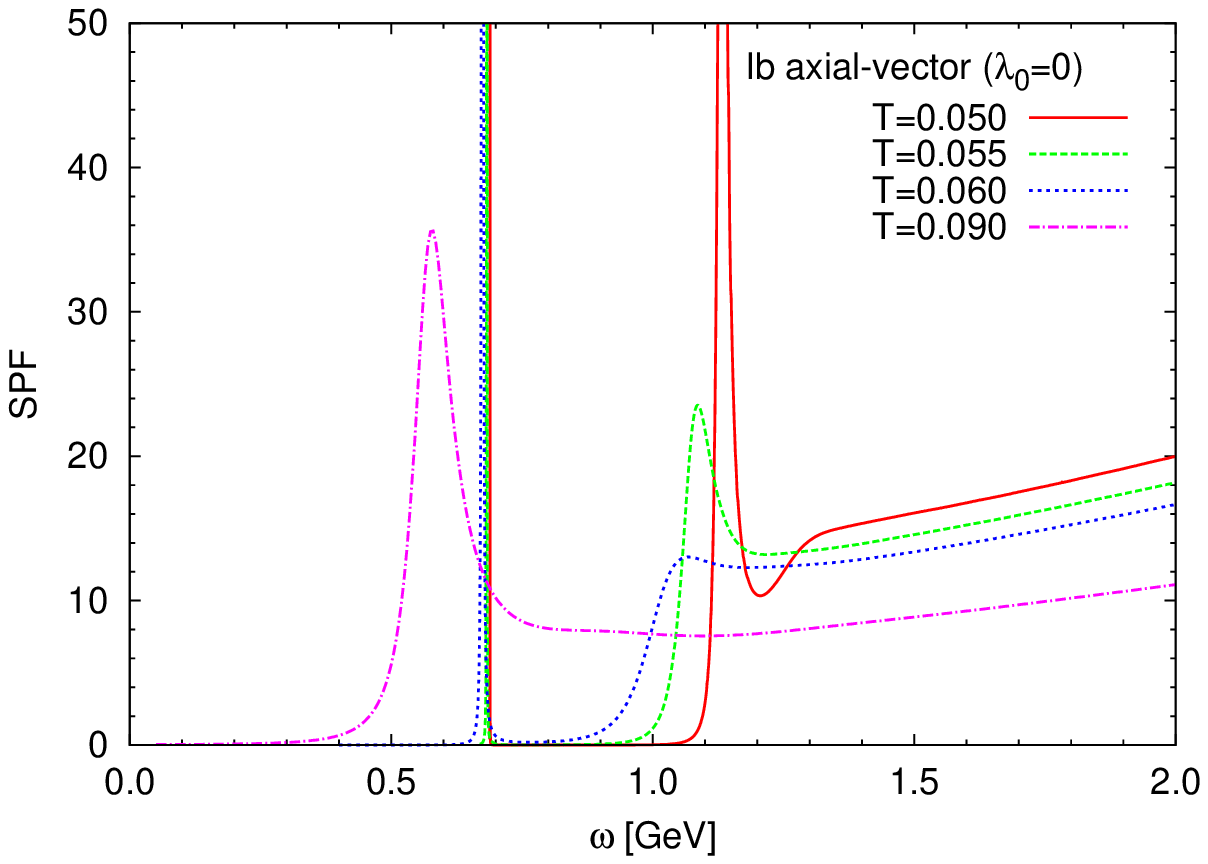}
\includegraphics[width=64mm,clip]{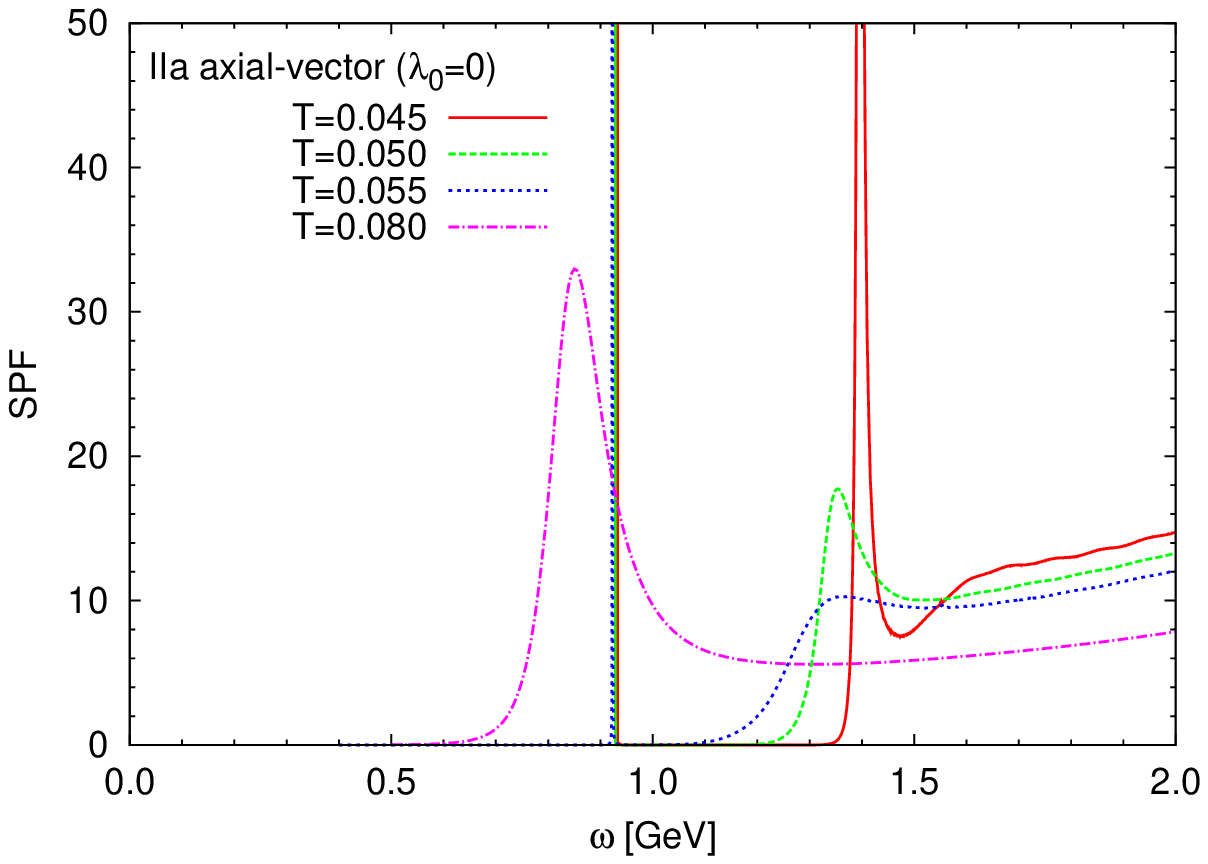}
\includegraphics[width=64mm,clip]{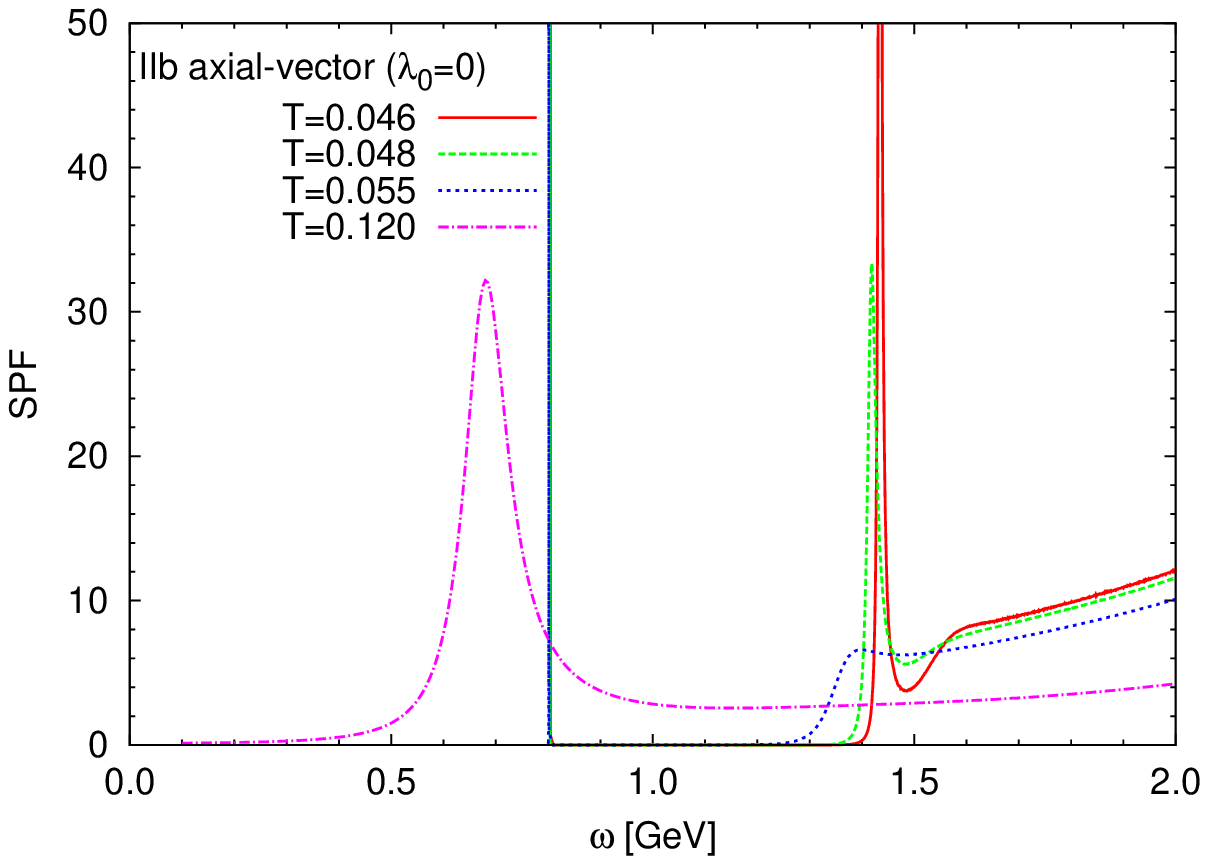}
\end{center}
\caption{
The results of the spectral function for axial-vector meson. Other choices are the same as ones in Fig.\ref{Fig_vec_nq}.
}\label{Fig_av_nq}
\end{figure}

\begin{figure}[!h]
\begin{center}
\includegraphics[width=64mm,clip]{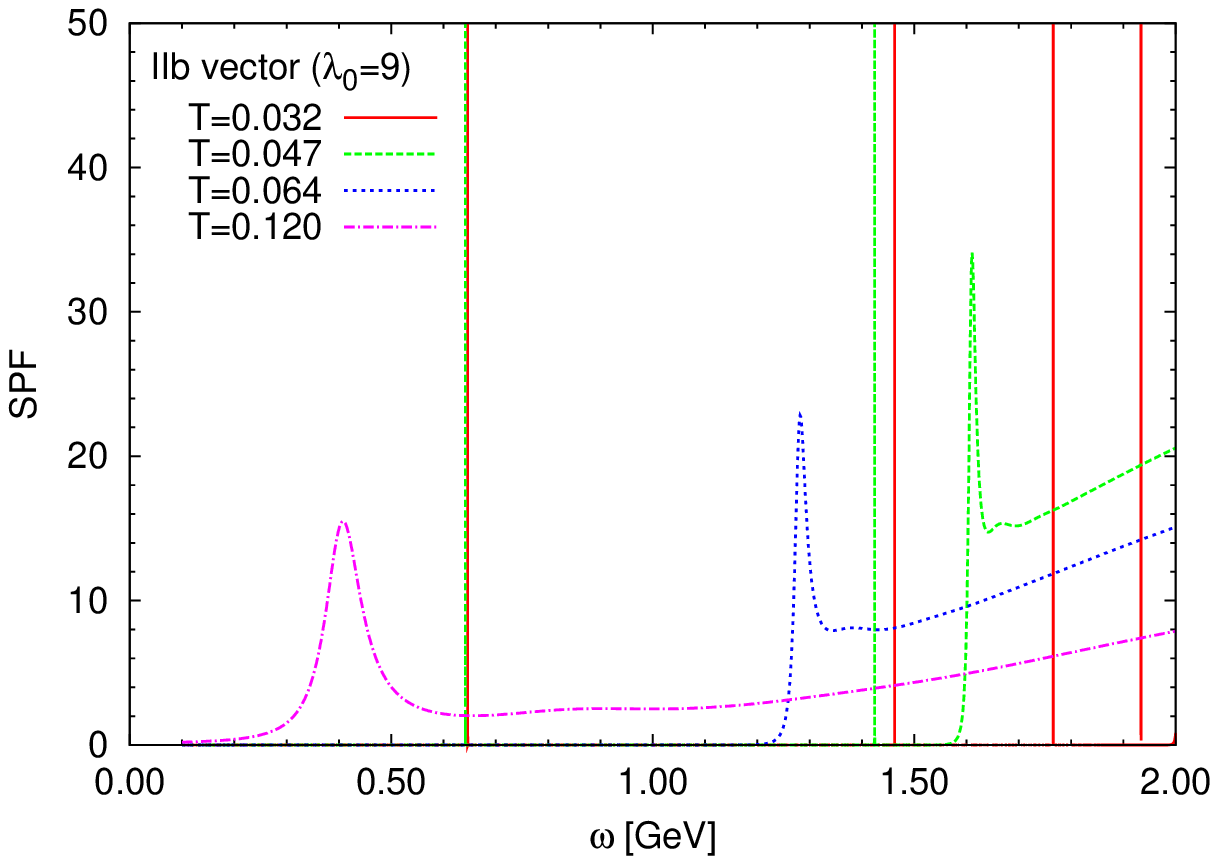}
\includegraphics[width=64mm,clip]{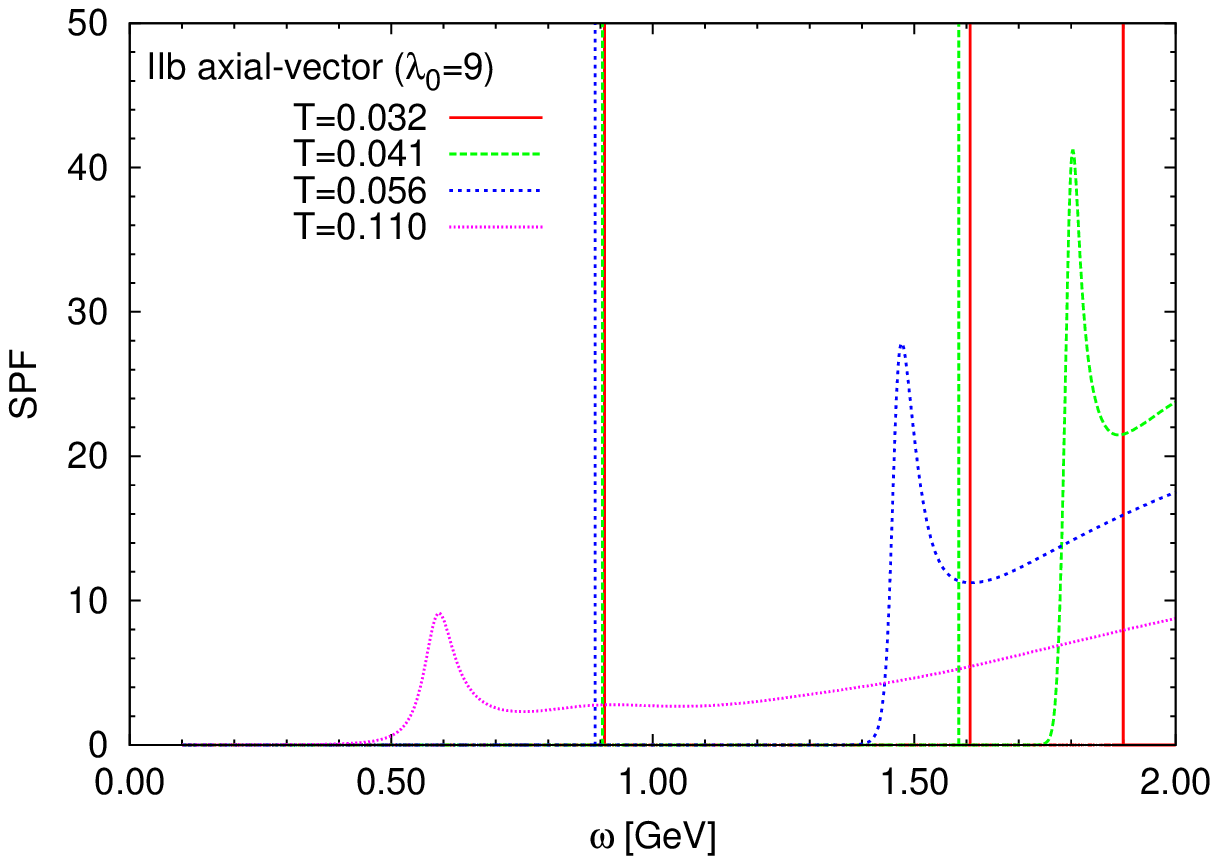}
%-----
\end{center}
\caption{
The results of the spectral function with the quartic term ($\lambda_0=9$).
Other choices are the same as ones in Fig.\ref{Fig_vec_nq} but with the parameters given Table.\ref{Tbl02w}.
}\label{Fig_va_qt}
\end{figure}

Now Let us compute the spectral function with $q=0$
and compare its result with the data listed
in Table.\ref{vecm_nqt} and \ref{vecm_qt}.
(As for $q \not = 0$, see Sec.\ref{Chap:MD_FME}.)
%---
The results are shown
in Fig.\ref{Fig_vec_nq}, \ref{Fig_av_nq} and \ref{Fig_va_qt}.
As we set $q=0$, the positions of the peaks in x-axis can read as the masses of mesons.

%=====================================
%
%
%=====================================
%
%
%=====================================
%
%
%=====================================

%=====================================
%
\begin{table}[!h]
\begin{center}
\begin{tabular}{c c c c c c | ccccc }
\hline\hline
  n & $\rho_{\rm V}$ (MeV) & Ia & Ib & IIa & IIb & $\rho_{\rm AV}$ (MeV) & Ia & Ib & IIa & IIb \\
  \hline %-------------------------
  0 & $775.5 \pm 1$ & 739  &   603 &   777 &   727 & 1230 $\pm$ 40      &  934 & 714  & 940  & 807\\
  1 & $1465 \pm 25$ & 1223 &  1175 &  1292 &  1468 & 1647 $\pm$ 22      & 1468 & 1247 & 1496 & 1507\\
  2 & $1720 \pm 20$ & 1534 &  1509 &  1596 &  1744 & $1930^{+30}_{-70}$ & 1822 & 1573 & 1831 & 1778\\
  3 & $1909 \pm 30$ & 1784 &  1769 &  1842 &  1971 & 2096 $\pm$ 122     & 2109 & 1829 & 2102 & 2003\\
  4 & $2149 \pm 17$ & 2000 &  1990 &  2054 &  2170 & $2270^{+55}_{-40}$ & 2358 & 2049 & 2338 & 2202\\
  5 & -             & 2193 &  2187 &  2249 &  2351 & -                  & 2582 & 2049 & 2338 & 2202\\
\hline \hline
\end{tabular}
\end{center}
\caption{The experimental and the predicted mass spectra for vector (left side) and axial-vector (right side) mesons without the quartic term ($\lambda_0=0$) given in \cite{pADSQCD1}.}
\label{vecm_nqt}
\end{table}
%
%=====================================
%
\begin{table}[!h]
\begin{center}
\begin{tabular}{ c c c c | c c c}
\hline\hline
  n & $\rho_{\rm V}$ (MeV) & IIa & IIb & $\rho_{\rm AV}$ (MeV) & IIa & IIb \\
  \hline
  0 & $775.5 \pm 1$ &  583 &   646 & 1230 $\pm$ 40      & 1128 &   913 \\
  1 & $1465 \pm 25$ &  900 &  1468 & 1647 $\pm$ 22      & 1643 &  1618 \\
  2 & $1720 \pm 20$ & 1248 &  1793 & $1930^{+30}_{-70}$ & 1953 &  1940 \\
  3 & $1909 \pm 30$ & 1564 &  2008 & 2096 $\pm$ 122     & 2225 &  2161 \\
  4 & $2149 \pm 17$ & 1860 &  2170 & $2270^{+55}_{-40}$ & 2486 &  2333 \\
  5 & $2265 \pm 40$ & 2143 &  2289 & -                  & 2742 &  2470 \\
\hline \hline
\end{tabular}
\end{center}
\caption{The experimental and the predicted  mass spectra for vector (left side) and axial-vector (right side) mesons with the quartic term ($\lambda_0=9$) given in \cite{pADSQCD1}.
}
\label{vecm_qt}
\end{table}
%
%=====================================

From the figures, it is seen that there are more spikes appearing at low temperature. In particular, the locations of the spikes which stand at lower temperature are very close to the values of the low-lying states obtained at zero temperature listed in Table.\ref{vecm_nqt} and \ref{vecm_qt}.
When the temperature is increased, the spikes become less and less and the peaks move to the low value of $\omega$.
It is noticed that the spikes collapse in turns from the higher excited states. Namely, the much higher a state is excited,
the more unstable that state becomes.

\begin{figure}[!h]
\begin{center}
\includegraphics[width=64mm,clip]{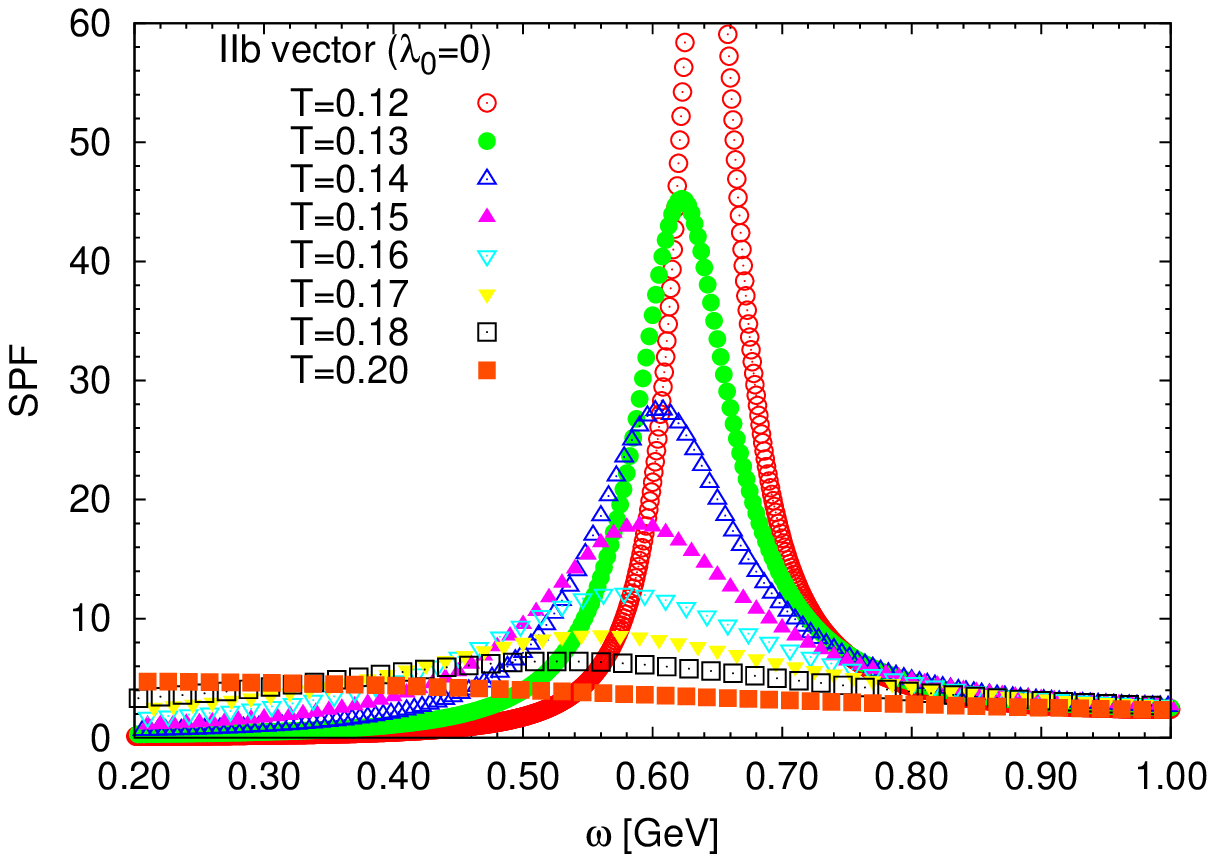}
\includegraphics[width=64mm,clip]{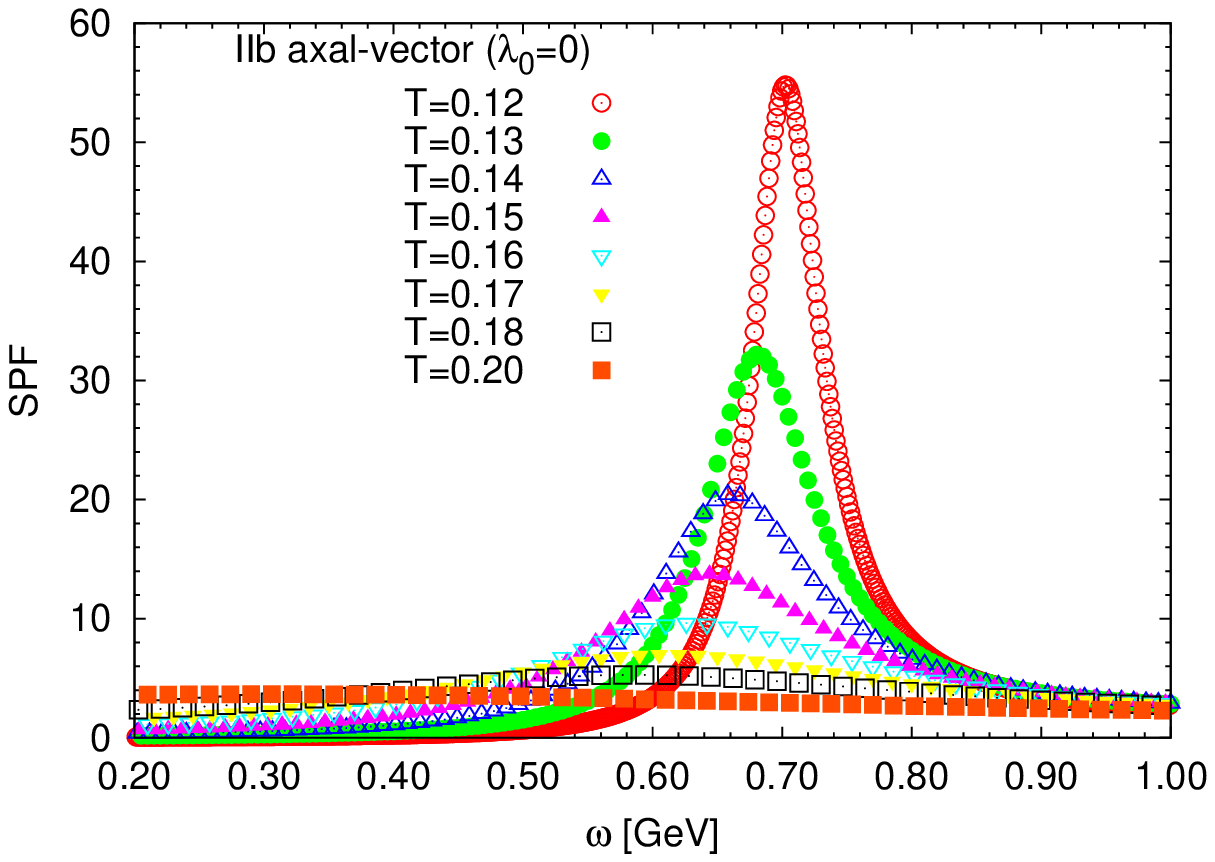}
\includegraphics[width=64mm,clip]{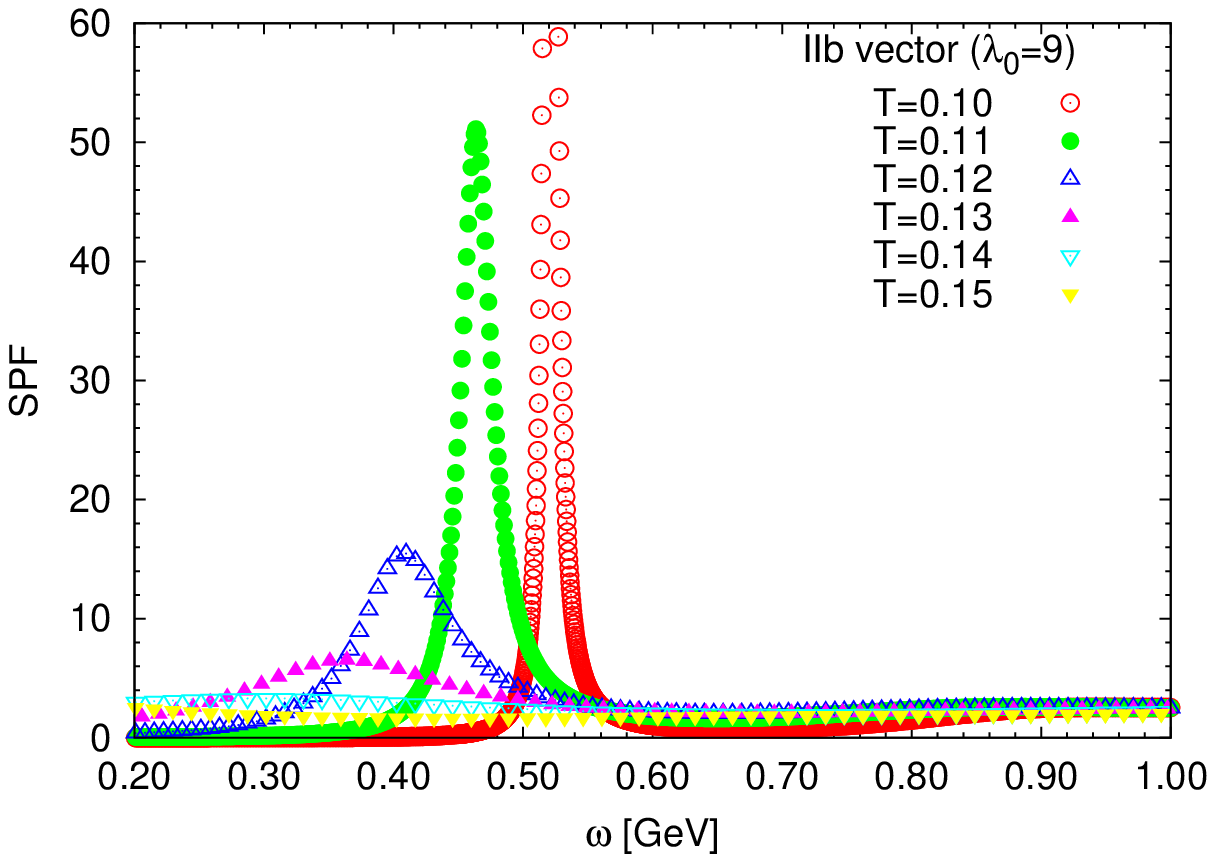}
\includegraphics[width=64mm,clip]{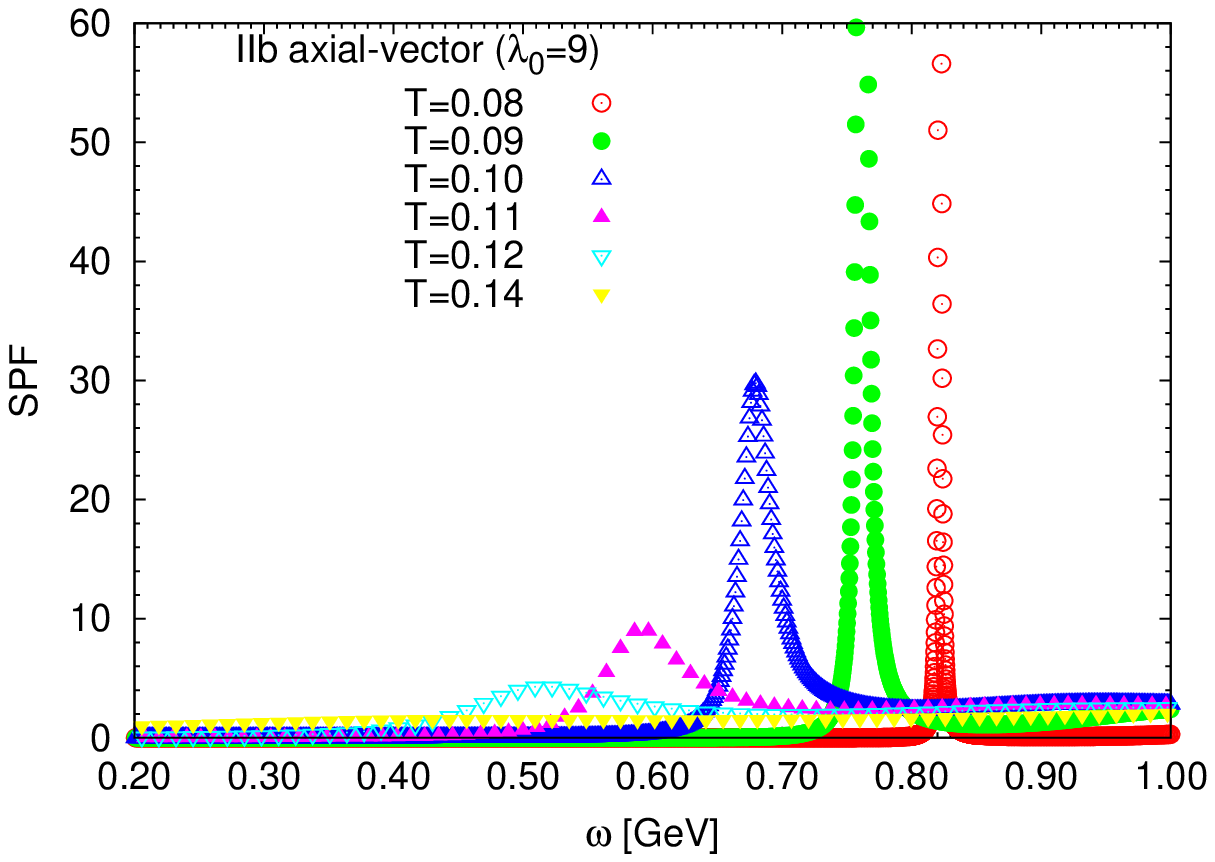}
\end{center}
\caption{
The peak dissolves as temperature grows for the lowest lying state.
The other choices are the same as ones in Fig.\ref{Fig_av_nq} or \ref{Fig_va_qt}.
}
\label{FigIIbHT}
\end{figure}

To check explicitly how the peaks behave when temperature becomes higher, we illustrate in detail the peak for the lowest lying state and plot in Fig.\ref{FigIIbHT} its behavior for various temperatures. It is seen that as the temperature becomes higher,
the width of the peak becomes wider and wider and eventually the peak disappears. We shall discuss it qualitatively in next section.

From the above studies,
we can draw the following statement that
the peaks standing like a spike
in the figures of the spectral function indicate the resonances of meson,
which can explicitly be seen
in Figs.\ref{Fig_vec_nq}, \ref{Fig_av_nq} and \ref{Fig_va_qt}
at low temperatures.
%---
When temperature increases,
the locations of the peak move
to a lower value of $\omega$
and the width of the peak becomes wider,
as explicitly seen in Fig.\ref{FigIIbHT}.
%---
Eventually, the peaks disappear completely at temperatures around $T=200$ MeV,
which is regarded as a critical temperature for the disappearance of mesons,
or for the confinement/deconfinement transition.

%---
%\textcolor{red}{
The results of Fig.\ref{FigIIbHT} imply that
the critical temperature
for the restoration of chiral symmetry breaking
in our model
should also locate around $T=200$ MeV. The reason is that
%---
the disappearance of the peak means
the deconfinement of quarks within the mesons. In a general consideration,
the restoration of chiral symmetry will occur
at the temperatures around that point.
On the other hand, the VEV $v(z)$ is going to become much small at the  temperatures around $T=200$ MeV .
This can be understood from the 5D AdS/QCD where the bulk coordinator $z$ plays the role of the running energy scale, i.e., a large $z$ corresponds to the low energy and a small $z$ to the high energy. The AdS/QCD is actually more reliable to describe the low energy dynamics of QCD with a large $z$, which may explicitly be seen from the wave functions of the mesons as shown in Fig.5 of Ref.\cite{pADSQCD1}, it is noticed that the peak of the wave function for the ground state appears at $z \sim 1.5$ GeV$^{-1}$ $\simeq 1/667$ MeV$^{-1}$, when $z < 1.0$ GeV$^{-1}$ the wave function goes far from the peak and becomes negligible small at $z < 0.5 $ GeV$^{-1}$, which corresponds to the energy scale around $\mu \simeq 2$ GeV where perturbative QCD gets available. Here, it is easy to check that at temperatures around $T=200$ MeV, the allowed value for $z$ is given by $z < 1/(\pi T) \simeq 1/628$ MeV$^{-1}$ due to the requirement for the black hole metric $f\geq 0$. From the boundary condition at UV, it is known that with the small values of $z$, the VEV $v_0(z)$ for $m_q =0$ approaches to zero as the power $z^3$ for the given condensation $\sigma$ at low energies, and the part $v_1(z)$ goes to zero exponentially. Thus it is expected that the VEV $v(z)$ holographically corresponding to the chiral condensation with the bulk coordinator $z$ characterizing the running energy scale approaches to be very small as $z$ goes to be at the UV boundary.
In this sense, we are led to the statement that the chiral symmetry will be restored around $T=200$ MeV.
%}

%\textcolor{red}{
Note that in ref. \cite{Misumi} the mass spectra in vector and axial-vector
have also been computed in a different model from ours and obtained a different dissolved temperature.
We shall make a comparison about the dissolved temperatures between two models and discuss the physical reasons causing the differences.
%-----
Let us first present the results for both models in Table.\ref{tbl_melt}.
Then it can clearly be seen that two results are quite different.
We now discuss the possible reasons causing the differences.
%}
One direct way to consider the difference is
to do an analysis for the existence of states as in Fig.1 in ref.\cite{Misumi},
here we will only make a general comment for the obvious differences in two kind of models.
There are actually three differences: the modified  background geometry due to $\mu_g^2$ as in eq.(\ref{f(z)}),
the dilaton and the bulk scalar field as in eq.(\ref{X(z)}) and (\ref{eomPhi2}),
and the quartic scalar field potential with the coupling $\lambda$ as in eq.(\ref{qtt}).
We shall make more detail comments as follows:

First, the introduction of the scale $\mu_g^2$ in the modified 5D metric will be crucial for causing the differences.
As the differences shown in Table.\ref{tbl_melt}
are the ones at finite temperature, and the scale $\mu_g^2$ couples with temperature in the metric, which may be seen from the redefined new variable $u = \pi T z$ so that the modified term  $\mu_g^2 z^2$ in 5D metric is related to the temperature via $\mu_g^2/(\pi^2 T^2) u^2$.
Thus tuning of $\mu_g^2$ term will show us how the effect of temperature is enhanced or softened.

Next, turning to the dilaton and the scalar field,
the dilaton in our model is determined as the solution of equation of motion with the given $v(z)$
which consists of the finite temperature and the zero-temperature parts
as in eq.(\ref{vassume}). It is different from the case in\cite{Misumi}
where the dilaton is fixed first and the VEV is determined from the equation of motion.
It is important to notice that to cooperate the chiral symmetry breaking and linear confinement
we have to obtain the correct boundary conditions for both the VEV at the UV boundary and the dilaton at the IR boundary,
for this purpose we have shown in\cite{pADSQCD1} that it is necessary have a non-zero $\mu^2_g$
or adding a high order interaction in the scalar potential.

Therefore, we arrive at conclusion that the deformation of IR region of bulk by the term $\mu_g^2$
is the key point for causing the differences given in Table.\ref{tbl_melt}.
%-----
Finally, it is clear from Table.\ref{tbl_melt} that
the effect of the quartic scalar field potential
with nonzero coupling $\lambda$ is the reason for lowering the critical temperature.
%}

\begin{table}[!h]
\begin{center}
\begin{tabular}{ c || c c | c c}
\hline\hline
                                   & $ \lambda_0=0$  & $\lambda_0=9$  & $ \lambda_0=0$ & $\lambda_0=9$  \\
\hline
Our results (IIb)                  & $0.20$           & $0.15$        & $0.20$         & $0.14$         \\
The results in ref.\cite{Misumi}   & $0.14$           & -             & $0.10$         & -              \\
\hline \hline
\end{tabular}
\end{center}
\caption{
This table shows the dissolved temperatures obtained
in vector (left) and axial-vector sectors (right)
in our present model and ref.\cite{Misumi}, where the unit is in GeV.
Here our results are obtained in model IIb.
As the quartic scalar potential is not taken into account in ref.\cite{Misumi},
their results correspond to $ \lambda_0=0$ case.
}
\label{tbl_melt}
\end{table}

Before closing this section, we would like to further make comments on the several options considered in this paper, which include:
the value for the coupling of the quartic scalar interaction
and the choices for the regularization parameters given in eq.(\ref{v1v}),
and the four types of models shown in table.\ref{Tbl01}. First, as for the coupling of the quartic scalar potential,
as mentioned in the introduction, it was first introduced for obtaining a nontrivial solution of dilaton and meanwhile yielding the chiral symmetry breaking, while it was pointed out that the bulk scalar potential field is not bounded due to the opposite sigen\cite{Gherghetta:2009ac}.
Concerning the values assigned in this study,
we have followed the analysis in ref.\cite{pADSQCD1} with changing the sign for the stability consideration and the nontrivial solution of dilaton is obtained as the consequence of the modified 5D metric at IR region with the introduction of the $\mu^2_g$ term, and the value of the coupling $\lambda$ is taken from ref.\cite{pADSQCD1} by fitting to the experimental data.
Concerning the regularization parameters $c_v$ and $c_{\lambda}$ as shown above, we have assigned the values
as small as possible in the numerical calculations, so that they do not affect our final conclusions.
As for the four types of models discussed originally in ref.\cite{pADSQCD1}, which is to show how the predicted mass spectra rely on the IR boundary behaviors and the different forms with the same boundary behavior of the VEV. As a consequence, it can be seen from ref.\cite{pADSQCD1} that the results are more sensitive to the boundary behaviors classified as 'a' and 'b' than the different forms named as 'I' and 'II' in Table.\ref{Tbl01},
the model IIb leads to more consistent results with experiment data than other types of models. This is the reason why we take the model IIb as the studying model in this paper.

\section{Mass shift and width of peak from spectral function}
\label{Chap:melt}

\begin{figure}[!h]
\begin{center}
\includegraphics[width=50mm,clip]{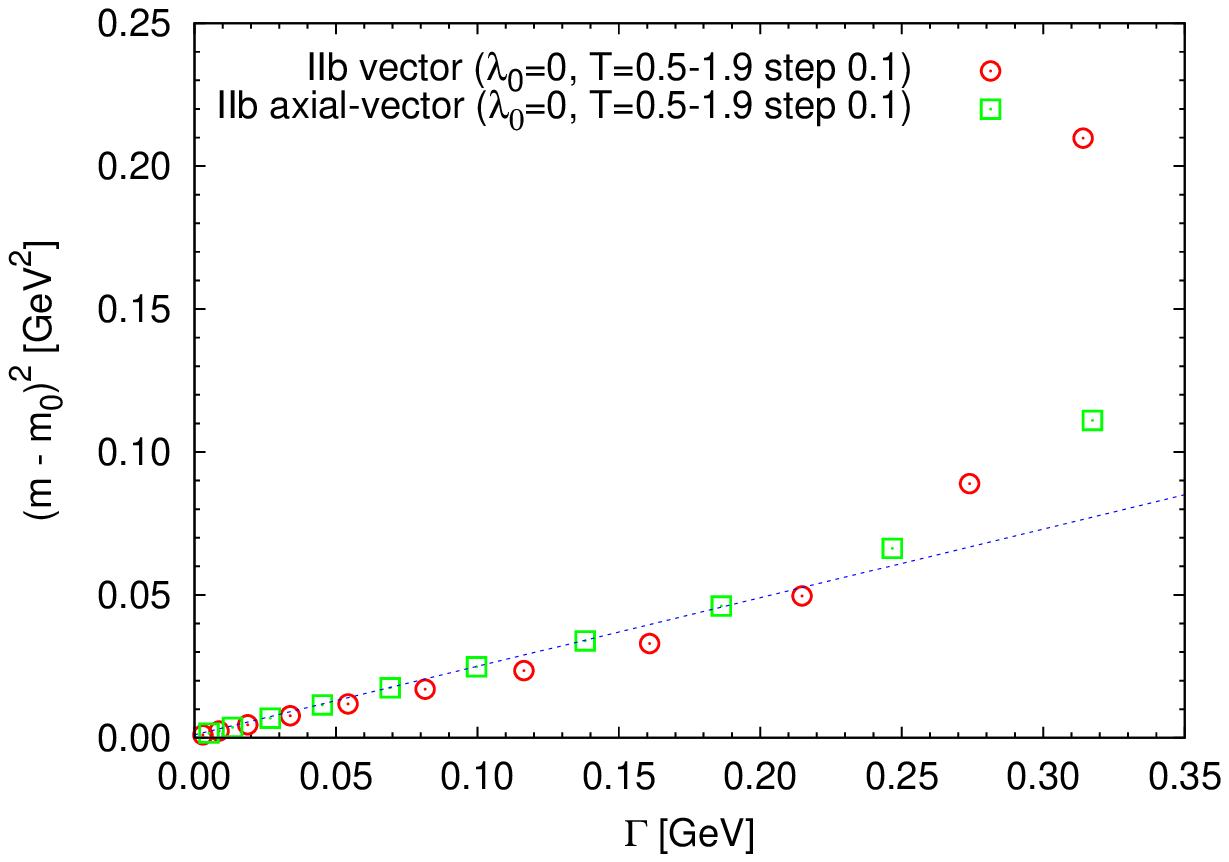}
\includegraphics[width=50mm,clip]{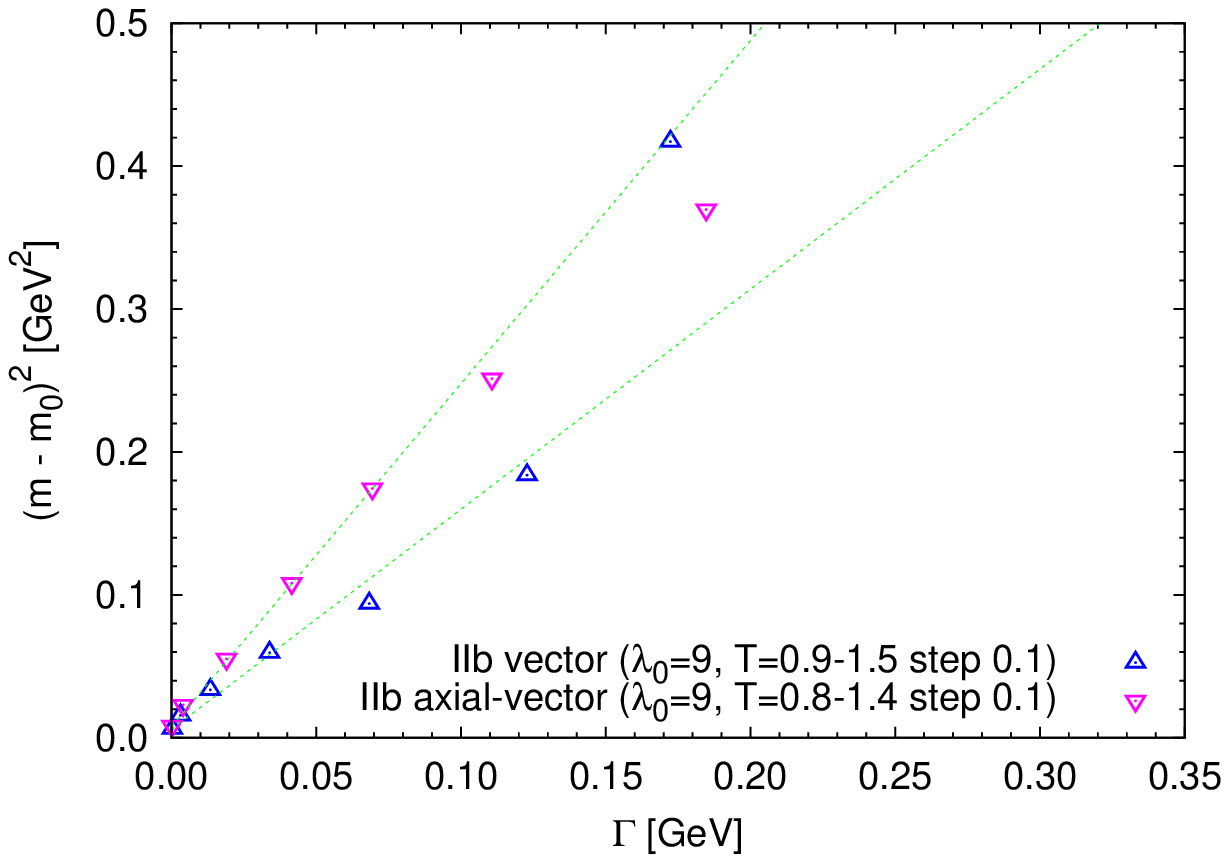}
\end{center}
\caption{
The relation between the mass shift $\Delta m^2 \equiv (m - m_0)^2$
and the width $\Gamma$, which is obtained from the fitting
to the numerical results of spectral function
for the lowest lying state by the Breit-Wigner form given in Eq.(\ref{fitfunc}).
The parameters used here are given in Eq.(\ref{v1v}) and Table.\ref{Tbl02wo} and \ref{Tbl02w},
and the unit of temperature is in GeV.
The dotted lines with the linear behavior are plotted for a comparison.
}
\label{FigMSIIb}
\end{figure}

\begin{figure}[!h]
\begin{center}
\includegraphics[width=50mm,clip]{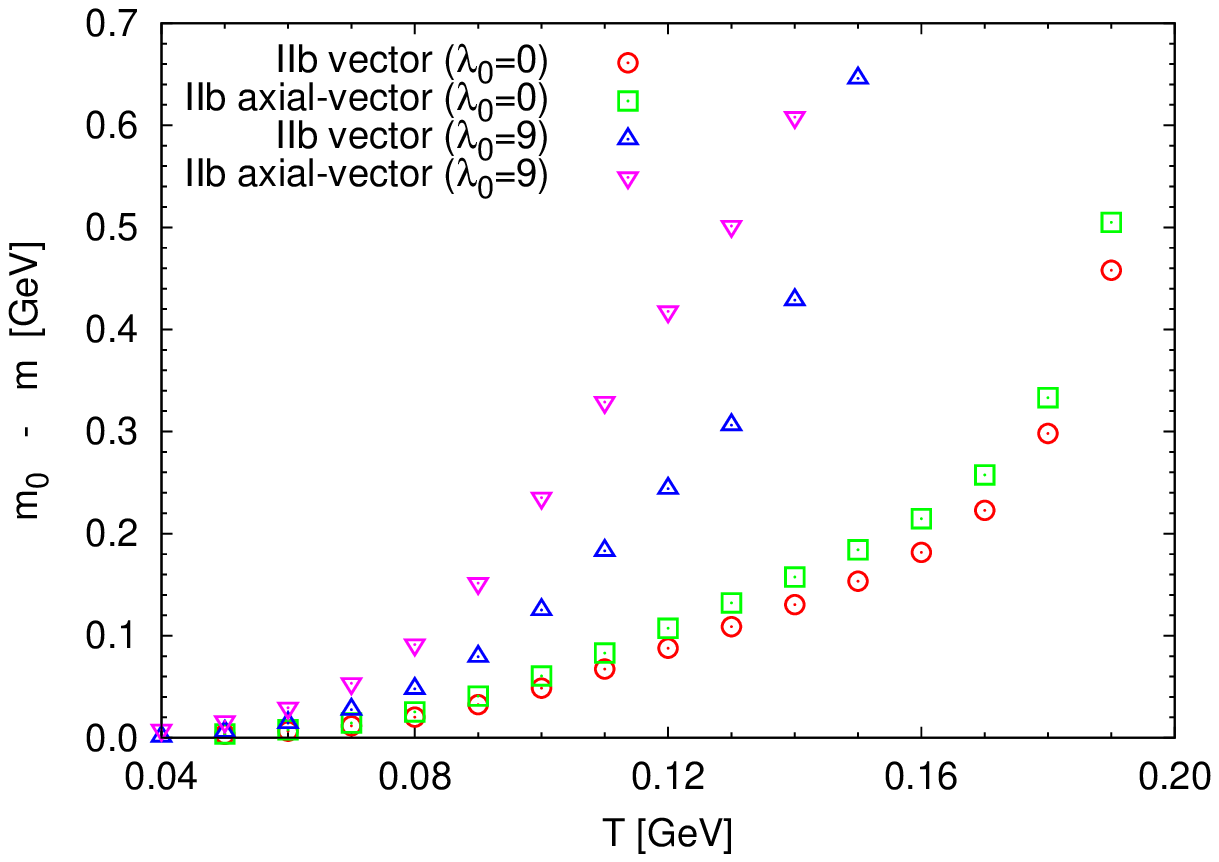}
\includegraphics[width=50mm,clip]{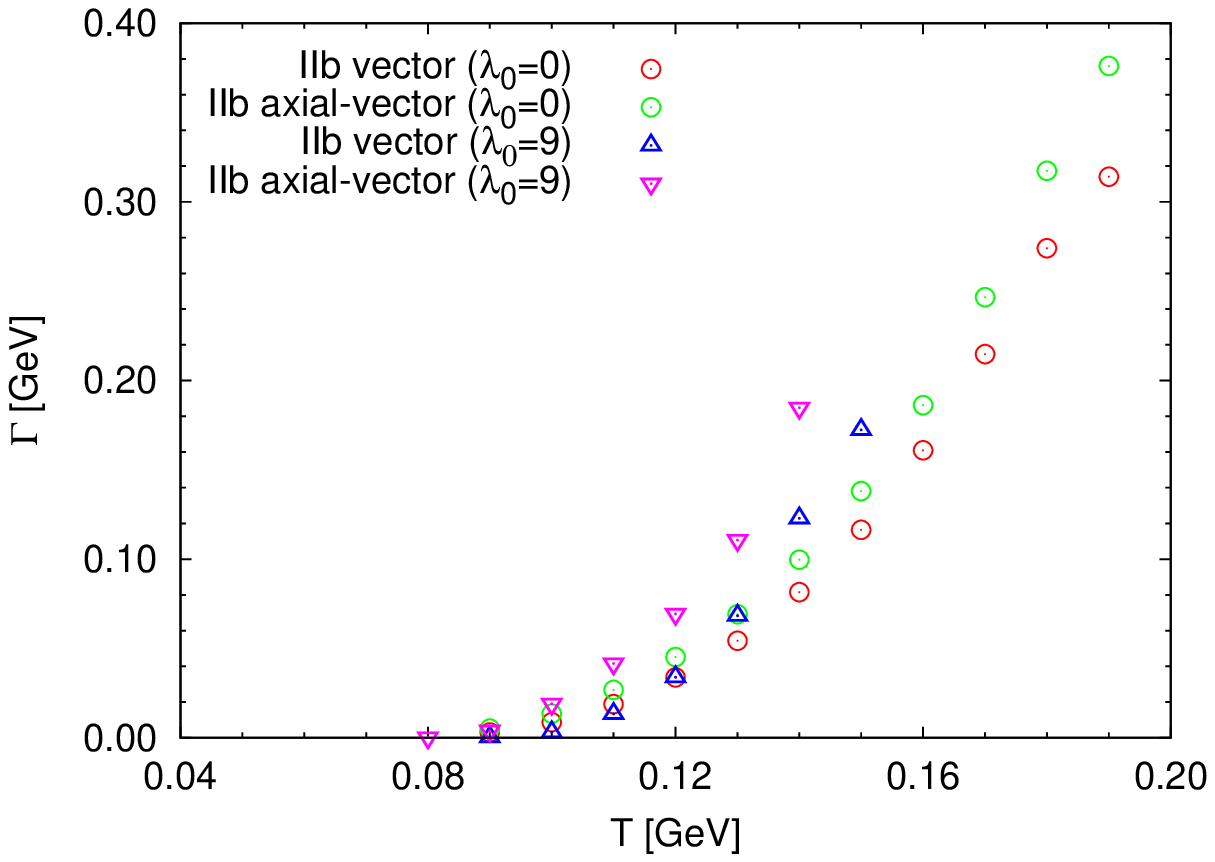}
\end{center}
\caption{
It shows the temperature effect in
$\Delta m \equiv m - m_0$ and $\Gamma$.
The other choices are the same as the ones in Fig.\ref{FigMSIIb}.
}
\label{FigMDTIIb}
\end{figure}

As we show in previous section how the peak dissolves as the temperature is increased to a critical point.
Here we are going to make a numerical study with fitting the spectral function by the following Breit-Wigner form:
\begin{equation}\label{fitfunc}
\frac{a \omega^b}{(\omega^2-m^2)^2+\Gamma^2} + P(\omega^2).
\end{equation}
Here $m$ and $\Gamma$ correspond to the location and width of the peak.
$P(\omega^2)$ represents the part not forming peaks and is taken to be the form  $P(\omega^2)=c_1 +c_2 \omega^2 +c_3 (\omega^2)^{c_4}$.
Thus it involves eight fitting parameters: $a,b,m, \Gamma,c_1,c_2,c_3$ and $c_4$.
To see quantitatively the mass shift and the width change of the peak, we consider the lowest lying state and plot the relation between the mass shift $\Delta m^2 \equiv (m-m_0)^2$ and the width $\Gamma$
against the temperature. Here $m_0$ is the predicted meson mass at zero temperature
given in Table.\ref{vecm_nqt} and \ref{vecm_qt}. The result is shown in Figs.\ref{FigMSIIb} and \ref{FigMDTIIb}.
It is seen from Fig.\ref{FigMSIIb} that
$\Delta m^2$ is proportional to $\Gamma$ as shown by the dotted line.
%---

From Fig.\ref{FigIIbHT}, it can explicitly be seen that the mass shift $\Delta m^2$ grows linearly as temperature increases
and begins to deviate from the linearity near the temperature which leads the spectral function to dissolve.
A similar phenomenology was also observed in ref.\cite{Misumi} and considered to be consistent
with the results based on QCD sum rule calculation\cite{QCDsumrule}.

It is noticed that the relation
between $\Delta m^2 \equiv (m-m_0)^2$ and $\Gamma$ is not known
in the low temperature region where a peak appears as a spike.
This is because there is no numerical results as in such a low temperature region,
the peak stands like delta function, it is difficult to measure the width numerically.
From Fig.\ref{FigMDTIIb}, it is seen that the linearity occurs above certain temperature. In the low temperature region, such a linearity changes since the mass $m$ and width $\Gamma$ get values $m=m_0$ and $\Gamma=0$ at zero temperature, respectively.
Thus, it is expected that a similar change happens in the relation
between $\Delta m^2 \equiv (m-m_0)^2$ and $\Gamma$.

\section{Finite momentum effect}
\label{Chap:MD_FME}

\begin{figure}[!h]
\begin{center}
\includegraphics[width=65mm,clip]{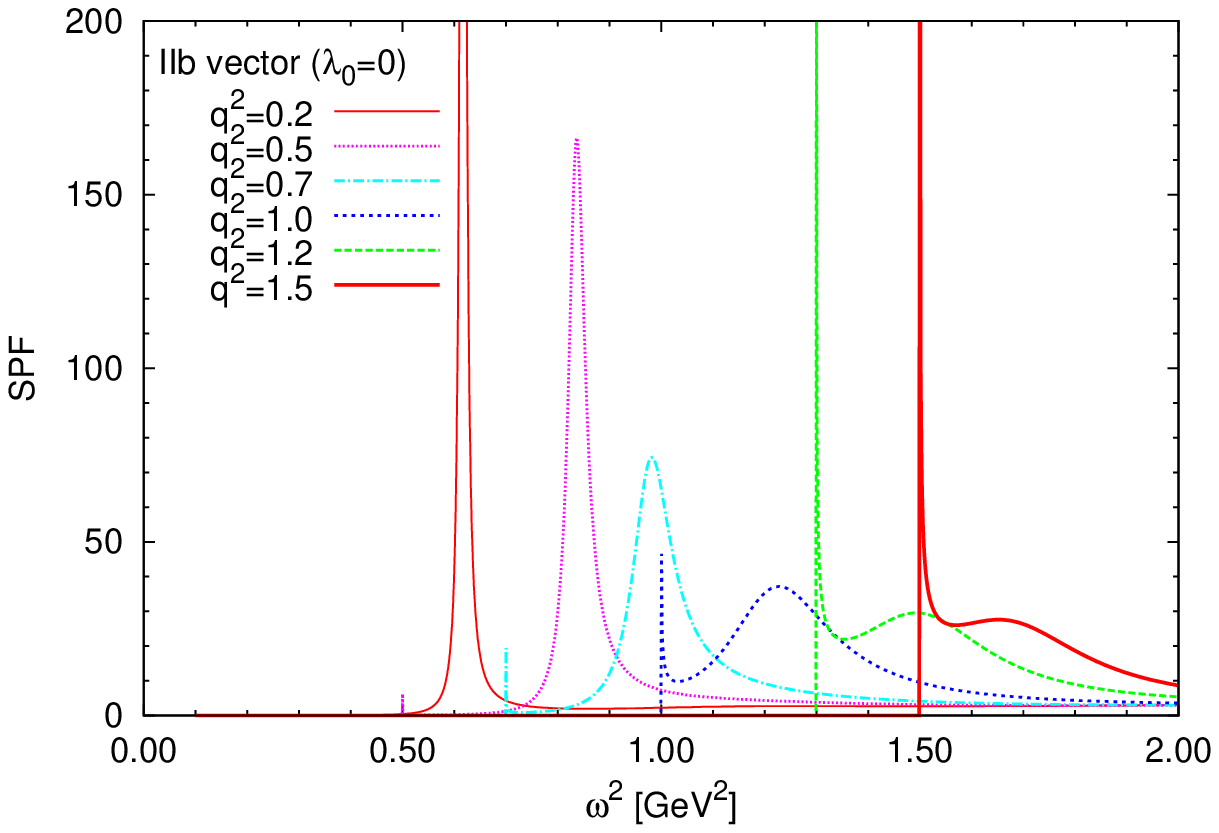}
\includegraphics[width=65mm,clip]{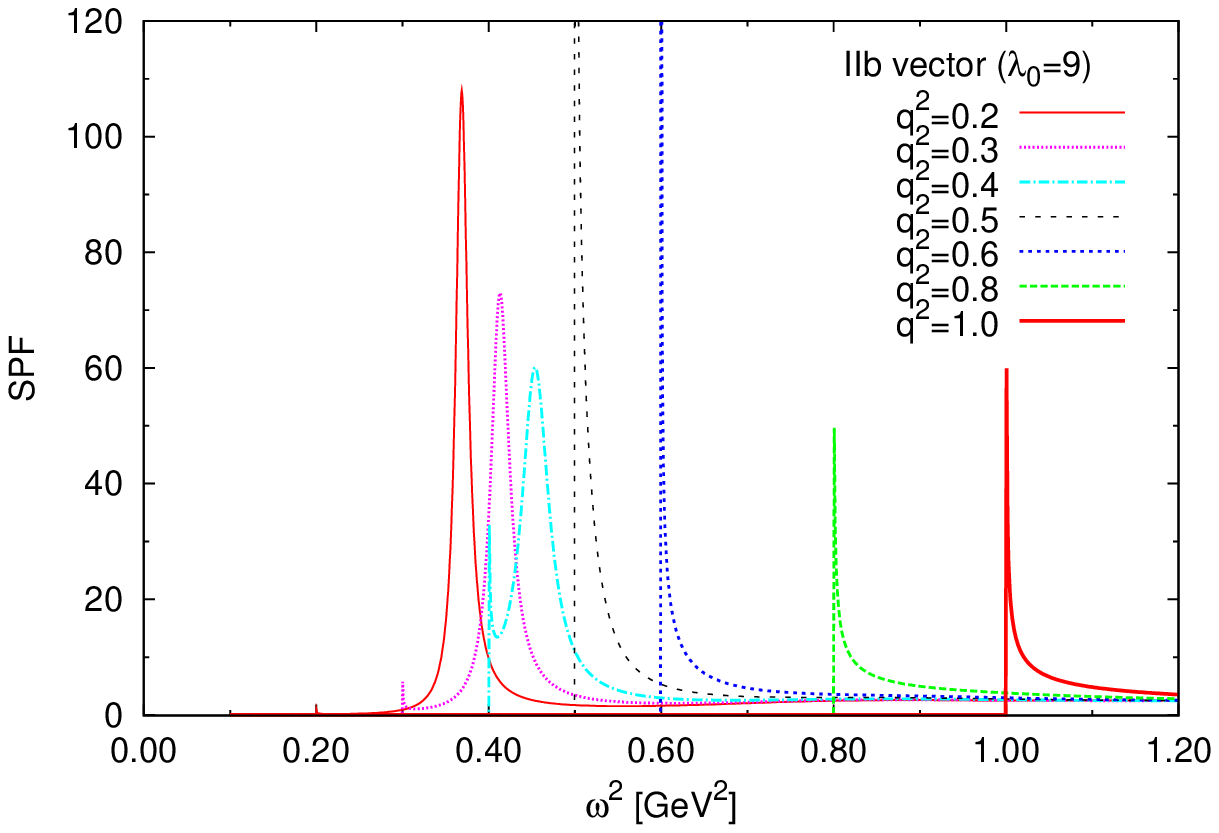}
\end{center}
\caption{
The momentum effect in the spectral function for various momentum in unit GeV with $T=0.10$ GeV.
The parameters in the left and the right are given in Eq.(\ref{v1v}), Table.\ref{Tbl02wo} and \ref{Tbl02w}.
The radial coordinate are taken to be $u \in [1-10^{-8},~10^{-8}]$ in the computation.}
\label{FigIIbFM}
\end{figure}

In this section, we are going to study the finite momentum effect in the spectral function. Such a momentum effect in the shape of spectral function was also studied in ref.\cite{Misumi,Colangelo:2009ra}. Our results are shown in Fig.\ref{FigIIbFM}.
Note that the spectral function in our present consideration is defined with a step function in Eq.(\ref{eq:spectralv})
for a positive squared energy in the boundary theory from a general point of view\cite{Kovst}.

It is seen that there is a wall-like behavior caused by the step function defined in Eq.(\ref{eq:spectralv}). In particular,
when the momentum turns on, the peak moves to large $\omega^2$ direction and gradually decays.
Such a phenomenology is consistent with the observation in \cite{Liu:2006nn,Faulkner:2008qk}, but different
from the results obtained by the perturvative analysis in a field theory\cite{Hidaka:2002xv}, which may be considered
as a peculiar and common behavior in strongly coupled field theories.

Let us turn to the effect of the step function defined in Eq.(\ref{eq:spectralv}).
Its effect appears as a wall in Fig.\ref{FigIIbFM}. It is seen that there is a blow-up around the wall,
and the hight of the wall is changed depending on fineness of step of $\omega^2$ taken in the numerical computation.
As a result for the introduction of the step function, one may notice the difference
in the speed of the wall moving as $q^2$ is increased. It may be seen from
the differences between the left plot and the right plot in Fig.\ref{FigIIbFM} with $\lambda_0=0$ and  $\lambda_0=9$,
as well as with parameters given in Table.\ref{Tbl02wo} and \ref{Tbl02w}.

\section{Conclusions and remarks}
\label{Chap:Sum}

In this paper,
we have extended the predictive soft-wall AdS/QCD model \cite{pADSQCD1,pADSQCD2}
to a thermodynamic model
by introducing a black hole metric. %to describe the finite temperature effect.
%---
As the predictive AdS/QCD model has adopted an IR-improved 5D metric
to characterize the low energy dynamics of QCD \cite{pADSQCD1,pADSQCD2},
we have shown that
it has to figure out a modified bulk vacuum expectation value
and a modified coefficient of quartic term
for the scalar field to get a non-singular dilaton solution.
%---
In obtaining the modified bulk vacuum expectation value
and a modified coefficient of quartic term,
an additional term is introduced to regulate the singular behavior.
For that,
the coefficients of the additional terms are taken
to be small as much as possible in the practical calculations.
With the smooth dilaton solution,
we have computed the spectral function
for the vector and axial-vector mesons.

It has been demonstrated that
the peaks in the spectral function indicate the resonance mesons,
as shown in Figs.\ref{Fig_vec_nq}, \ref{Fig_av_nq} and \ref{Fig_va_qt}.
%---
The location and the width of the peaks
moves to a lower value of $\omega$ and becomes wider
as the temperature increases, respectively, as seen in Fig.\ref{FigIIbHT}.
%---
We have found that the peak disappears completely
at temperatures around $T=200$ MeV,
which implies the deconfinement of quark
and the restoration of chiral symmetry breaking.
%----
To see how the peak dissolves quantitatively
when the temperature is increasing to the critical point,
we have made a numerical study
by fitting the spectral function with taking the Breit-Wigner form.
%----
It has been seen that the mass shift $\Delta m^2=(m-m_0)^2$ with $m_0$ as the meson mass at zero temperature
grows in a linear way
as temperature increases.
%----
We have also investigated momentum effects and shown that
the peak moves to a large $\omega^2$ direction.
%-----

\vspace{1 cm}

\centerline{{\bf Acknowledgement}}
\vspace{20 pt}

One of authors (S.T.) would like to thank
the administrators in the cluster system in National Technical University of Athens
and B-Factory Computer System of KEK and Azuma Takehiro
for his technical tips in manipulating of the clusters,
as well as to Kenji Fukushima and Tatsuhiro Misumi.
We would like to thank the referee for the useful comments.
This work was supported in part by the
National Science Foundation of China (NSFC) under Grant \#No.
10821504, 10975170 and the key project of the Chinese Academy of Science.

%%%%%%%%%%%%%%%%%

\appendix

%%============================================
\section{The formula for two-point retarded Green function}
\label{App:HSF}
%%============================================

In this appendix, we will show how to obtain Eq.(\ref{BoA}).
It is seen that the Lagrangian up to third-order perturbations of vector and axial-vector sector can be written as
\begin{equation}\label{app01}
S_{\rm gauge}
=
-\int \! dz~
\frac{e^{-\Phi(z)}a(z)}{2 (g_5)^2}
\int \! d^4p~ e^{i p \cdot x} \left\{
f(z)
\big( \widetilde{V}^a_x{}'(p,z) \big)^2
+
\frac{\omega^2}{f(z)}
\big( \widetilde{V}^a_x(p,z) \big)^2
\right\} +\cdots,
\end{equation}
where ``$\cdots$'' denotes the axial-vector part which has the same form as the vector part and we abbreviate it.

In general, the action is assumed to have the following form
\begin{equation}\label{gene1}
S[\phi] = \int \! dz~
\Big\{
{\cal A}(z) \phi(z)\phi''(z) + {\cal B}(z) (\phi'(z))^2 + {\cal C}(z) \phi(z)\phi'(z) + {\cal D}(z) (\phi(z))^2
\Big\},
\end{equation}
where $\phi(z)$, ${\cal A}(z)$, ${\cal B}(z)$, ${\cal C}(z)$ and ${\cal D}(z)$ are arbitrary coefficients.
It can be shown that Eq.(\ref{gene1}) may be rewritten as follows
\begin{equation}\label{gene2}
S[\phi] =
\frac{1}{2}
\int \! dz~
\Big\{ {\rm EOM} \cdot \phi(z) + \partial_z ({\rm ST}) \Big\}.
\end{equation}
``EOM'' denotes equations of motion and it will vanish as the bulk gravity is evaluated at classical level.
On the other hand, ``ST'' denotes a surface term given as
\begin{equation}
{\rm ST} \equiv
{\cal A}(z) \phi(z)\phi'(z)
+ \big\{ \phi(z){\cal A}(z) - \phi(z){\cal A}'(z) \big\}\phi(z)
+ 2 {\cal B}(z)\phi(z)\phi'(z)
+ {\cal C}(z)\phi^2(z).
\end{equation}
Here we will mention  briefly how to obtain eq.(\ref{gene2}) from eq.(\ref{gene1}):
Firstly, we take the variation in an usual way to get equations of motion as
$S[\phi +\delta \phi]-S[\phi]$.
Then it can be summed up in ``EOM'' part and ``ST'' part up to the second order, i.e.,
$\displaystyle
S[\phi + \delta \phi]-S[\phi] =
\int \! dz~
\Big\{ {\rm EOM} \cdot \delta\phi(z) + \partial_z ({\rm ST}) \Big\} +{\cal O}( \delta \phi^2)$.
Then, discarding the higher-order terms and replacing $\delta \phi$ with $\phi$,
its lhs becomes $2S[\phi]$ (provided that powers of all fields are same), and we arrive at Eq.(\ref{gene2}).

Using this prescription,
eq.(\ref{app01}) can be written at classical level as
\begin{equation}
S_{\rm gauge}
=
\left.
-\frac{e^{-\Phi(z)}a(z)f(z)}{2(g_5)^2}
\int \! d^4p~ e^{i p \cdot x}
\widetilde{V}^a_x(p,z) \widetilde{V}^a_x{}'(p,z) \right|^1_{z=0} +\cdots.
\end{equation}
The evaluation at $z=0$ in the above can be written as
\begin{eqnarray}
&&
\left.
\frac{e^{-\Phi(z)}a(z)f(z)}{2(g_5)^2}
\int \! d^4p~ e^{i p \cdot x}
\widetilde{V}^a_x(p,z) \widetilde{V}^a_x{}'(p,z) \right|_{z=0} \nonumber \\
&=&
\left.
\int \! d^4p~ e^{i p \cdot x}
\frac{\widetilde{V}^a_x(p,z) \widetilde{V}^a_x{}'(p,z) }{2(g_5)^2z}\right|_{z=0}\nonumber \\
&=&
\left.
\int \! d^4p~ e^{i p \cdot x}
\frac{A^a(\omega,q)A^a(-\omega,-q)}{g_5^2}
\left[
\frac{B^a(\omega,q)}{A^a(\omega,q)}
+
\frac{q^2 - \omega^2}{2}
\left\{
\gamma_E + \log \left( \frac{z}{2} \sqrt{ \omega^2-q^2} \right)
\right\}
\right]\right|_{z=0}.
\end{eqnarray}
Here we have substituted eq.(\ref{eq:Asy45}).
$A^a(\omega,q) = A^a(-\omega,-q)$ as can be seen from eq.(\ref{eq:eomV1}) and (\ref{eq:eomAV2}).
$\gamma_E$ means Euler constant.
Thus following GKP-W relation and the prescription given in  ref.\cite{Son:2002sd},
the retarded Green function
$ \displaystyle
G^a(\omega,q) =
-i \int \! d^4x~e^{-q \cdot x} \theta(t)
\big\langle \big[J^a_x(x), J^a_x (0)\big]\big\rangle
$ (As for what $J^a_x(x)$ means, see below eq.(\ref{def_VW}).) can be obtained as
\begin{equation}
G^a(\omega,q)= -
\frac{2}{g_5^2}
\left[
\frac{B^a(\omega,q)}{A^a(\omega,q)}
+
\frac{q^2 - \omega^2}{2}
\left\{
\gamma_E + \log \left( \frac{z}{2} \sqrt{ \omega^2 - q^2} \right)
\right\}
\right]\Bigg|_{z=0}.
\end{equation}
Thus from the imaginary part given in $B^a(\omega,q)/A^a(\omega,q)$, we obtain the spectral function for vector and axial-vector mesons.

\end{document}